\documentclass[twocolumn]{article}
\usepackage[margin=.5in]{geometry}
\usepackage{mathpazo}
\usepackage{makecell}
\usepackage{bm}

\usepackage{float}
\usepackage{lipsum}

\usepackage{comment}
\usepackage{makecell}
\usepackage{colortbl}
\usepackage[utf8]{inputenc}
\usepackage{microtype}
\usepackage{xcolor,pifont}
\usepackage[T1]{fontenc}

\usepackage{tikz}
\usepackage{pgfplots}
\pgfplotsset{compat=1.18}

\usepackage{microtype}
\usepackage{graphicx}
\usepackage{booktabs} 
\usepackage[hidelinks]{hyperref}

\usepackage{amsmath}
\usepackage{amssymb}
\usepackage{mathtools}
\usepackage{amsthm}

\usepackage{multirow}

\usepackage{algorithm}         
\usepackage[noend]{algpseudocode}     
\usepackage{xspace}
\usepackage{subcaption}

\usepackage{bm}

\theoremstyle{plain}

\theoremstyle{definition}

\theoremstyle{remark}

\usepackage{xcolor}
\usepackage{soul}

\usepackage{color}
\definecolor{light}{rgb}{0.5, 0.5, 0.5}

\usepackage{xcolor}
\usepackage{soul}

\title{Kernpiler: Compiler Optimization for Quantum Hamiltonian Simulation with Partial Trotterization}

\author{
    Ethan Decker\footnote{Corresponding author. \texttt{ecd5249@upenn.edu}} \\
    University of Pennsylvania
    \and
    Lucas Goetz \\
    ETH Zurich
    \and
    Evan McKinney \\
    University of Pittsburgh
    \and
    Erik Gustafson \\
    (RIACS) at NASA Ames Research Center
    \and
    Junyu Zhou \\
    University of Pennsylvania
    \and
    Yuhao Liu \\
    University of Pennsylvania
    \and
    Alex K.\ Jones \\
    Syracuse University
    \and
    Ang Li \\
    Pacific Northwest National Laboratory
    \and
    Alexander Schuckert \\
    University of Maryland
    \and
    Samuel Stein \\
    Pacific Northwest National Laboratory
    \and
    Eleanor Crane \\
    Massachusetts Institute of Technology
    \and
    Gushu Li \\
    University of Pennsylvania
}

\begin{document}

\maketitle

\begin{abstract}
Quantum computing promises transformative impacts in simulating Hamiltonian dynamics, essential for studying physical systems inaccessible by classical computing. However, existing compilation techniques for Hamiltonian simulation — in particular, the commonly used Trotter formulas — struggle to provide gate counts feasible on current quantum computers for beyond-classical simulations. We propose partial Trotterization, where sets of non-commuting Hamiltonian terms are directly compiled, allowing for less error per Trotter step and therefore a reduction of Trotter steps overall. Furthermore, a suite of novel optimizations is introduced which complement the new partial Trotterization technique, including reinforcement learning for complex unitary decompositions and high-level Hamiltonian analysis for unitary reduction.  We demonstrate with numerical simulations across spin and fermionic Hamiltonians that compared to state-of-the-art methods such as Qiskit's Rustiq and Qiskit's Paulievolutiongate, our novel compiler presents up to $10\times$ gate and depth count reductions.
\end{abstract}


\section{Introduction}

\begin{figure*}
    \centering
    \includegraphics[width=.95\linewidth, height=5.5cm]{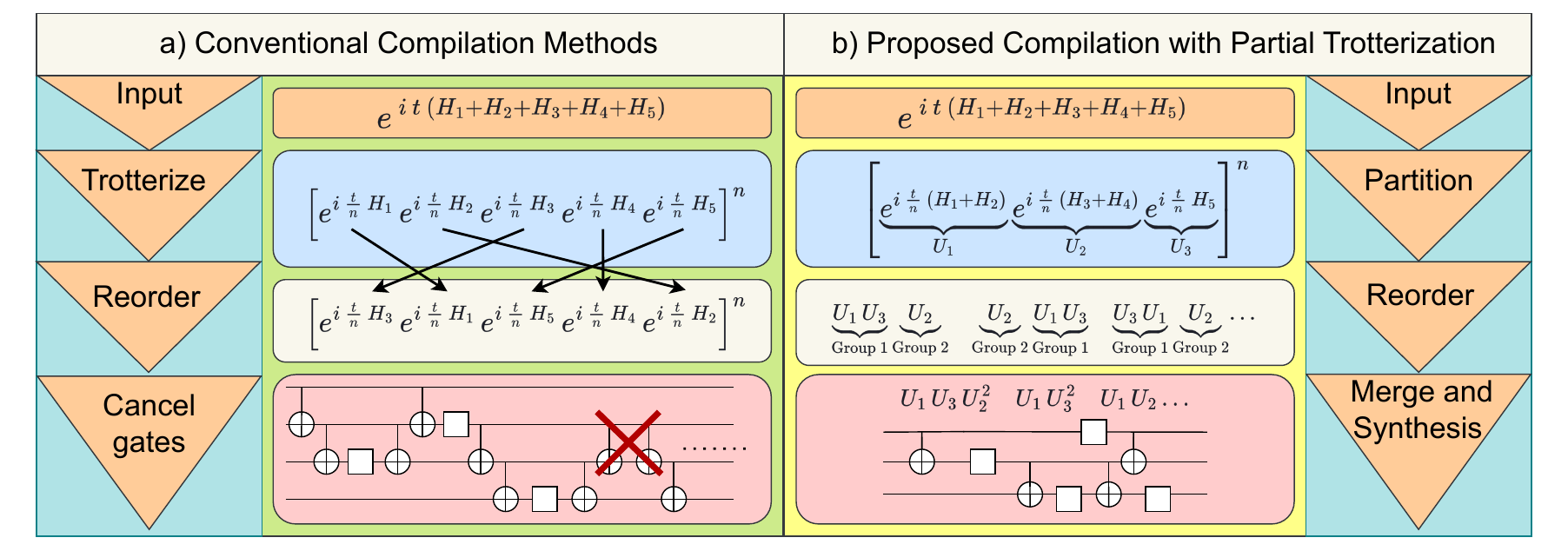}
    \caption{\textbf{Conventional compilation flow vs the proposed Kernpiler compiler}. b) Pipeline for reducing gates through error term reduction. First we group into partial Trotter steps which act on a subset of N qubits, in our case N=3. Then we perform an efficient numerical rewrite of the partial Trotter unitaries. Next step, group into commuting subsets of unitaries placing the largest two groups of unitaries on the edges of the Trotter step. Finally, we use a partially symmetric Trotter step to cancel error terms in the expansion by alternating every other Trotter steps order. Commuting unitaries then merge back together naturally allowing for a unitary reduction with no additional error. The compilation finishes at circuit-level.
    }
    \label{fig:pipelineoverview}
\end{figure*}
Quantum computing holds immense promise as a paradigm-shifting technology, with one of its most impactful applications lying in \emph{Hamiltonian simulation}~\cite{lloyd1996universal,Childs2012,low2019,Childs_2021}---the process of evolving a qubit array according to the physics (Hamiltonian) of a target quantum system. Hamiltonian simulation is widely recognized as a cornerstone of quantum computing’s value proposition, as it enables the study of complex physical phenomena that elude classical methods, promising advances in materials science~\cite{babbush_low-depth_2018}, quantum chemistry~\cite{Cao_2019}, nuclear-~\cite{Bauer:2023qgm} and high-energy physics~\cite{crane2024}. However, bringing these benefits to fruition requires efficient compilation strategies to convert the Hamiltonian time evolution to the quantum gate sequences.

Existing efforts in quantum simulation compilation, beyond higher-level compilers such as~\cite{Stavenger:2022wzz, mcclean2019openfermionelectronicstructurepackage}, have employed the domain knowledge and Pauli algebra to optimize the quantum Hamiltonian simulation circuit.
In the conventional compilation flow for quantum Hamiltonian simulation (on the left of Figure~\ref{fig:pipelineoverview}), a Hamiltonian, $H$, will first be decomposed into a sum of weighted terms, e.g.  Pauli strings, $H=\sum_i H_i$ (weights absorbed to $H_i$'s). The Trotter product formula then allows one to approximate the Hamiltonian time evolution $e^{iHt}$ with a long sequence composed of each individual Hamiltonian term, $e^{iH_it}$, for time evolution.
Existing optimization approaches include simultaneous diagonalization of commuting Pauli strings in the decomposition~\cite{Cowtan_2020, cowtan2020genericcompilationstrategyunitary, van_den_Berg_2020}, Pauli string reordering optimizations after Trotterization~\cite{li2021paulihedralgeneralizedblockwisecompiler, gui2021termgroupingtravellingsalesperson, anastasiou2022tetrisadaptvqe}, Pauli network synthesis\cite{debrugiere2024fastershortersynthesishamiltonian, paykin2023pcoastpaulibasedquantumcircuit}, etc., which have yielded noticeable benefits.


The drawback of such conventional compilation flow for quantum Hamiltonian simulation is that each of the Hamiltonian terms must individually be decomposed into its own unitary. 
Therefore, all these compilation approaches rely on the vanilla error bound in the Trotter formula~\cite{Hatano_2005} and focus on reducing the number of gates per Trotter step. Furthermore, to reduce the approximation error one must then increase the number of Trotter steps according to this bound. 
Notably, for spin and fermionic Hamiltonians, achieving high fidelity with low approximation error typically demands extraordinarily long quantum circuits \cite{Campbell_2019, Childs_2021,hemery_measuring_2024}. 

The objective of this paper is to show a new path forward for quantum Hamiltonian simulation by incorporating the optimization opportunity from error analysis. We observe that the fundamental bottleneck of product formulas arises from \emph{error scaling}, wherein non-commuting Hamiltonian terms are approximated by sequential exponentials.
As the error in Trotterization is directly dependent on the non-commutativity of Hamiltonian terms, strategies to mitigate this characteristic in a fine-grained manner can provide a new and scalable way for continued progress in Hamiltonian simulation.


To this end, we propose the new paradigm of \textit{Partial Trotterization} for Hamiltonian compilation, as depicted on the right side of Figure~\ref{fig:pipelineoverview}. Along with this novel concept, we develop a suite of optimizations, namely Kernpiler, which complement partial Trotterization to command large reductions over modern full Trotterization techniques.  
\textbf{First,} rather than fully decomposing each Hamiltonian term as a separate exponential, we partially Trotter the input Hamiltonian by partitioning non-commuting Hamiltonian terms together into more complex unitaries. We then manipulate and decompose multi-term exponentials instead of exponentials of individual terms. 
This can significantly improve the error scaling compared with conventional full Trotterization.
\textbf{Second,} after the partial Trotterization, our Kernpiler groups commuting unitaries together and orders the exponentials of the partially Trotterized Hamiltonian terms to maximize the gate cancellation and term merging. The terms within each group are shuffled at every Trotter step to avoid systematic approximation errors. 
\textbf{Third,} at the final stage, we propose a Monte Carlo Tree Search (MCTS) method to synthesize the exponential of partially Trotterized Hamiltonian terms into a highly optimized basic gate sequence. To maintain the search efficiency, we only search for coupling structures in the MCTS framework, while the single-qubit gates are realized via differentiable methods. 
This allows us to fully exploit the potential of error reduction from partial Trotterization.


Theoretical analysis shows that Partial Trotterization can effectively lower the Trotter depth (and thus the gate count) needed to reach a desired accuracy, yielding a quadratic reduction in circuit depth as a function of group size for first- and higher-order Trotterization. 
We also conduct numerical simulation for a range of benchmark Hamiltonians (Heisenberg, Ising, Fermi--Hubbard, etc.) with diverse localities, geometries, and term weights.
The results show that Kernpiler outperforms Qiskit's Rustiq~\cite{debrugiere2024fastershortersynthesishamiltonian} and Qiskit's Paulievolutiongate (Paulihedral)~\cite{li2021paulihedralgeneralizedblockwisecompiler} with up to a 86\% (40\% on average) reduction in depth and CNOT gate count along with up to a 85\% (11\% on average) reduction in single qubit gates (comparing against whichever does better between Rustiq and Paulihedral).

Our major contributions can be summarized as follows:
\begin{enumerate}
    \item We propose a new decomposition technique, Partial Trotterization, for reducing the error per Trotter step in product formulas. 
    \item We propose a series of compilation algorithms , Kernpiler,  to group the Hamiltonian terms, reorder and merge the grouped Hamiltonian terms, and synthesize the exponential of the grouped terms into basic gates. 
    \item Experimental results show that Kernpiler outperforms Qiskit's Rustiq~\cite{debrugiere2024fastershortersynthesishamiltonian} and Qiskit's Paulievolutiongate~\cite{li2021paulihedralgeneralizedblockwisecompiler} with significant gate count and circuit depth reduction.
\end{enumerate}

\section{Background} 

In this section, we introduce the necessary background to understand the proposed optimization on quantum Hamiltonian simulation. 
For basic quantum computing concepts (e.g., qubit, gate, linear operator, circuit), we recommend~\cite{nielsen2010quantum} for more details.

\subsection{Hamiltonian Simulation, Pauli Strings, and Trotterization}

The time evolution of a quantum system with its Hamiltonian $H$ is characterized by the operator $e^{iHt}$ where $t\in\mathbb{R}$ representing the time.
In general, directly translating the $e^{iHt}$ into a quantum algorithm is hard and a principled approaches are required. 

In this passage we introduce the concept of Pauli string and Hamiltonian decomposition.
In an \(n\)-qubit system, a Pauli string is defined as a length-\(n\) tensor product of the operators \(\{X, Y, Z, I\}\), where each operator acts on a specific qubit index. This direct mapping of Pauli strings to qubits naturally arises in many quantum Hamiltonians, making them a convenient basis for both theoretical analyses and practical implementations.

The time evolution of a Pauli string, \(P\) is $e^{iPt}$ and it can be synthesized into a quantum circuit using a series of Pauli gates, CNOT gates, and a  Z-rotation gate exactly. This process works straightforwardly when dealing with a single Pauli string; however, challenges emerge when the objective is to synthesize an exponential of a sum of Pauli strings, \(\exp\!\bigl(it\sum_iP_i\bigr)\). In these cases, closed-form analytical decompositions generally do not exist, which motivates the use of approximation techniques to break down the weighted sum of Pauli Strings into implementable quantum gate sequences.

It is known that all Pauli strings of length $n$ formulate a basis for the linear space of all the Hermitian operators over $n$-qubits, and Hamiltonians are Hermitian operators.
So a Hamiltonian can always be decomposed into a weighted sum of Pauli strings $H=\sum_iw_iP_i$ where $w_i\in\mathbb{R}$.
For simplicity, we absorb the weight and the associated Pauli string into one Hamiltonian term and denote $H=\sum_iH_i$ in the rest of this paper.
To approximate the exponential of the sum of Hamiltonian terms, one commonly employs \emph{Trotterization}. Formally, it is based on the Lie--Trotter formula \cite{Hatano_2005}:
\begin{align}
\label{commutation_error}
e^{t(H_i+H_j)} &\approx \left(e^{\frac{t}{N}H_i}\, e^{\frac{t}{N}H_j}\right)^N,\\
\left\| e^{t(H_i+H_j)} - \left(e^{\frac{t}{N}H_i}\, e^{\frac{t}{N}H_j}\right)^N \right\| &\le \frac{t^2}{2N}\,\|[H_i,H_j]\| + \mathcal{O}\!\Bigl(\frac{t^3}{N^2}\Bigr),\notag
\end{align}
where $N$ is the number of Trotter steps, and the error depends on the sum of commutators \([H_i, H_j]\) mitigated linearly by the number of Trotter steps. By splitting a large sum into smaller components that can be individually exponentiated, Trotterization provides a systematic method for approximating time-evolution operators. Increasing the number of Trotter steps reduces the approximation error but also increases the overall circuit depth. 
This method has been implemented in many industry and academia-offered software development kits \cite{qiskit2024,cirq_developers_2024_11398048,Killoran_2019} as a standard approach for quantum Hamiltonian simulation. 

\subsection{Randomized Compilation} 

Randomized compilation has recently gained considerable attention in the quantum computing community as a means to mitigate coherent errors in quantum circuits. 
By converting systematic error into stochastic error, randomized compilation can improve the robustness of quantum algorithm approximations by allowing for better asymptotic scaling on larger time simulations. 
Early theoretical frameworks for randomized compilation were first presented in \cite{Campbell_2019,Winick:2022scr,2016efficienttwirling,PhysRevA.94.052325,2018efficienttwirling,2013PhRvA..88a2314G}, illustrating how randomly selected gate layers can effectively reduce correlated noise processes. 
In product formulas, random compilation can be invoked by shuffling each Trotter step, which would then cause rapidly changing signals and evolutions to average out erroneous terms \cite{Childs_2019}, to give better scaling. 
This work leverages the idea of randomization to shuffle the orderings of partially Trotterized terms (introduced later) to turn coherent error into stochastic error. 

\subsection{Reinforcement Learning Algorithms and Monte Carlo Tree Search}
\begin{figure}[t]
    \centering
    \includegraphics[width=1\linewidth]{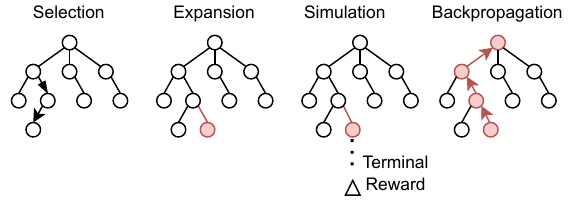}
\caption{The four stages of the Monte Carlo search tree. 1. Selection of a node for expansion and evaluation. 2) Expansion: choosing a new action and state combination that has not been explored. 3) Simulation: Randomly traversing states and actions to a terminal state and evaluating the outcome. 4) Backpropagation: updating tree metadata on outcomes learned through simulation}
    \label{fig:mctsintro}
\end{figure}
In this paper, we will also use a reinforcement learning framework to synthesize some unitary operators into basic gates. Here we briefly introduce the framework of the Monte Carlo Tree Search (MCTS) algorithm.

When the structure of a problem is only partially known or highly complex, \emph{reinforcement learning (RL)} offers a powerful framework for decision-making and optimization. It balances the fundamental trade-off between \emph{exploration}---searching for new strategies---and \emph{exploitation}---refining known, successful strategies. Within RL, MCTS is a well-established technique that represents a system in terms of states and actions. 
To decide which states are valuable and which actions to take to reach valuable states, RL algorithms employ a policy. 
A policy describes how the algorithm interacts with the environment and is learned over many iterations or attempts. 

An MCTS utilizes a tree data structure where actions are represented by edges and states by nodes. The algorithm is fundamentally a Markovian process, where the next action taken is only dependent on the current state. 
By balancing exploration and exploitation appropriately, our traversal policy should converge to an accurate representation of the value of being in any individual state and therefore allow for a more optimal selection of states and actions over greedy or dynamic programming based approaches. 

MCTS proceeds in four key phases (see Figure. \ref{fig:mctsintro}):

\begin{enumerate}
    \item \textbf{Selection.} From the root of the search tree, MCTS traverses down to leaf nodes following a policy that balances visiting promising states with exploring unvisited ones.
    \item \textbf{Extension.} At an unvisited leaf, any unexplored actions lead to new states. MCTS selects an action from the leaf and adds the resulting state to the tree.
    \item \textbf{Simulation.} To quickly estimate the value of this newly added state, MCTS conducts a \emph{Simulation}---a rapid simulation or heuristic-based approximation---until reaching a terminal condition.
    \item \textbf{Backpropagation.} The outcome of the simulation is then propagated back up the tree to update value estimates and guide future searches.
\end{enumerate}

This iterative process of selection, extension, simulation, and backpropagation allows MCTS to allocate computational effort to promising areas of the solution space while maintaining coverage of unexplored regions.

\section{Opportunities and Challenges} \label{sec:opporutnities}

\textbf{Opportunity}: Our optimization opportunities come from fine-grained analysis of the error terms in the approximation. 
The error between the Trotter product formula and exact Hamiltonian time evolution can be shown through the BCH formula \cite{Hatano_2005}. The formula states: 
\begin{equation}
\log\Bigl(e^{\Delta t H_i} e^{\Delta t H_j}\Bigr)
= \Delta t\, H_i + \Delta t\, H_j
+ \frac{(\Delta t)^2}{2}[H_i,H_j] 
+ \cdots
\end{equation}

When approximating $
\log\bigl(e^{\Delta t(H_i+H_j)}\bigr) 
$ with $ \Delta t H_i + \Delta tH_j 
$, the dominant error term is $(\Delta t)^2 [H_i,H_j] + \cdots$. 
 The higher-order nested commutators are of order $ (\Delta t)^3$ and beyond.
 The primary optimization opportunity identified in this work is to reduce the effect of these commutators.
As a small example, consider the following Hamiltonian with 4 terms where none commute with each other:\[
\begin{aligned}
H &= H_i + H_j + H_k + H_l, \ \text{where}\\
H_i &= X_1 Y_2 Z_3, \ H_j = Y_1 Z_2 X_3&\\
H_k &= Z_1 X_2 Y_3, \ H_l = X_1 Z_2 X_3.&
\end{aligned}
\]

Now, naive Trotterization would give an error of the form: 
\begin{align}
\epsilon_{\text{full Trotter}} \propto &[H_i,H_j] + [H_i,H_k] + [H_i,H_l] + [H_j,H_k]\notag\\
&+ [H_j,H_l] + [H_k,H_l]
\end{align}

However, if we did not fully Trotterize the Hamiltonian and instead kept $H_i+H_j$ and $H_k+H_l$ in the exponentials (see Figure~\ref{fig:pipelineoverview}), there would be a smaller bound on the error term: 
\[
\epsilon_{\text{partial Trotter}} \propto [H_i,H_k] + [H_i,H_l] + [H_j,H_k] + [H_j,H_l]
\]
This motivates us to consider grouping terms to contract the additive errors that arise from Trotterization.
By strategically partitioning non-commuting operators into commuting partitions, we can potentially reduce the commutator error between terms, leading to lower overall Trotterization error and step counts. 
However, partitioning the Hamiltonian terms will immediately bring two challenges listed as follows.


\textbf{Challenge 1:}
The first question is how we can partition the terms effectively.
The objective of partitioning the Hamiltonian terms is to let the partitions be as dense as possible so that the follow-up compilation has more potential to rewrite the circuit with more gate count reduction. 
Without dense partitions, our rewrites would be very similar to the naive CNOT tree decomposition of the Hamiltonian simulation compilation due to the lack of opportunity for gate cancellations in the rewrite. 
Existing quantum program partitioning mostly focus on gate-level circuit partitioning for circuit resynthesis~\cite{Daei_2020}, \cite{kaur2025optimizedquantumcircuitpartitioning} which only collects adjacent gates. 
Other partitionings for specific Hamiltonians have also been explored \cite{Mart_nez_Mart_nez_2023}, however, existing partitioning techniques have not been generalized to other Hamiltonians of interest, and often require pre-processing circuits to allow for partitions to be analytically decomposed.
Therefore, we believe that there is improvement to be made for Hamiltonian partitioning on the axes of generality and efficiency.

\textbf{Challenge 2:}
Suppose we make a partition of Hamiltonian terms $H_i$, $H_j$, and $H_k$. The second challenge is how to efficiently compile and optimize the unitary $e^{it(H_i+H_j+H_k)}$ as there is no established approach for the complicated exponentials.
Previous approaches mostly focused on implementing the exponential of individual terms~\cite{li2021paulihedralgeneralizedblockwisecompiler}, \cite{debrugiere2024fastershortersynthesishamiltonian}, \cite{Kalajdzievski_2018}.
If we implement the exponential of these terms one by one, we naturally resort to the vanilla Trotterization and lose all the benefits of error reduction from partitioning. 
Additionally, there exists general unitary decompositions \cite{PhysRevApplied.22.034019}, \cite{Shende_2006}, however the gate counts of these methods are very high and can hurt complexity savings from the partitions. 
Consequently, exploring more efficient approaches for decomposing unitaries is motivated by the hypothesis that using high level Hamiltonian structure and learning algorithms will allow for more efficient circuits. 

We now summarize the opportunities and challenges.
For conventional full Trotterization, the error at each step is relatively high, leading to a high Trotter step count while implementing the circuit of the exponential of individual Hamiltonian terms is easy.
On the other hand, 
implementing partial Trotterization by partitioning the Hamiltonian terms will reduce the error and thus yield a low Trotter step count while the lack of efficient unitary decomposition methods may negate gates saved through less steps. 
Overall, our objective 
is to use partial Trotterization with a new term partitioning method and a new unitary decomposition method for the exponential of many Hamiltonian terms, achieving low Trotterization step count and low gate count in unitary decomposition simultaneously.

\section{Kernpiler Framework} 

In this section, we introduce in detail the Kernpiler framework that can deeply optimize the quantum Hamiltonian simulation by leveraging the optimization opportunities and overcoming the challenges mentioned above. 

\subsection{Overview}

The Kernpiler framework is outlined in Figure.~\ref{fig:pipelineoverview}b). The input is a quantum Hamiltonian for which the user wishes to obtain $e^{iH t}$ for a set time $t$. 

Firstly, the input is partially Trotterized. 
For example, instead of fully Trotterizing $e^{i (H_1 + H_2 + H_3)t}$ to $e^{i H_1 t}e^{i H_2 t}e^{i H_3 t}$, the algorithm may partially Trotterize to $e^{i (H_1 + H_2)t}e^{i H_3 t}$. 
To do this, partitions must be formed by sorting Hamiltonian terms based on their operator weight (e.g., $X_1 X_2 X_3$ which acts on three qubits is a weight $3$ term), constraining each partition to not act on more than $n$ qubits, where $n$ can be chosen arbitrarily. 
This results in the dense unitaries labeled $U_i$ in Figure.~\ref{fig:pipelineoverview}b). 
Because, in order to do this, the entire circuit needs to be searched, this is the Challenge 1 which we referred to as dense circuit partitioning as discussed in section~\ref{sec:opporutnities}, and which we solve by remaining at a higher level operator representation, referred to as high-level circuit partitioning. 
Later, these $n$ weight unitaries will be decomposed directly using reinforcement learning methods. Because decomposing arbitrarily high weight unitaries is hard, in the rest of this paper we choose $n=3$, however we will also comment on choosing larger $n$ later.

Secondly, the partially Trotterized unitaries are grouped such that in each group, the unitaries commute. 
After constructing groups of commuting unitaries, the order of groups within the Trotter step is determined. 
For our implementation, two groups containing the most and second most unitaries are placed on the edge of the Trotter step. 
In every step the side in which the two groups are placed is flipped such that neighboring Trotter steps have at their adjacent edges the identical commuting groups (these will be merged in the following step). 

Thirdly, still at the Hamiltonian term level, adjacent identical groups which commute, (i.e., $[U_i,U_j]=0$) are merged together (i.e., $U_i U_j U_j U_i$ is `merged' to $U_i^2 U_j^2$). 
After merging groups, there will still be a source of error that comes from the non-commuting terms within a single Trotter step (see Eq.~\ref{commutation_error}). 
This approximation error would be repeated each time the Trotter step is applied. We refer to this as coherent noise. 
To counteract this, we randomly shuffle the order of the terms within each successive Trotter step maintaining terms in their respective groups such that this noise becomes stochastic.
The k
Kernpiler framework then concludes with rewriting the dense unitaries into a target gate set to be executed on a quantum computer. 


\subsection{Hamiltonian Partitioning Algorithm}

\begin{table}[t]
    \centering
        \caption{Input is an array of Pauli strings. First the algorithm sorts the array on the highest qubit indices acted upon with tiebreakers being the weight of the string. Next the terms are grouped in a greedy fashion such that in each group the terms act on no more than 3 unique qubit indices.}
    \renewcommand{\arraystretch}{1.5} 
    \begin{tabular}{|c|c|}
        \hline
        \textbf{Step} & \textbf{Terms} \\ \hline
        Input & $[X_3, \quad X_1X_2, \quad Z_3Z_4, \quad Z_1]$ \\ \hline
        Sort & $[Z_1, \quad X_1X_2, \quad X_3, \quad Z_3Z_4]$ \\ \hline
        Group & $[Z_1, \quad X_1X_2] , [X_3, \quad Z_3Z_4]$
        \\ \hline
        Result & $e^{i \frac{t}{n} (Z_1 + X_1 X_2)} = U_1, \quad e^{i \frac{t}{n} (X_3 + Z_3 Z_4)}=U_2$ \\ \hline
    \end{tabular}

    \label{fig: greedypatition}
\end{table}



The first stage in our compilation pipeline is the partitioning step (shown in Table~\ref{fig: greedypatition}), which allocates Pauli strings into partitions for partial Trotterization. 
The goal is to maximize the density of terms which do not commute in each partition.
The input to this Figure is an array of Hamiltonian Pauli terms and the output is partitioned sets of Hamiltonian terms.
Currently, each partition of Hamiltonian terms can act non-trivially on 3 qubits maximum, and the unitary made from the partitioned Hamiltonian terms needs to be of size $8\times8$. 
Different from circuit-level partitioning strategies, which can only partition a few adjacent gates \cite{Daei_2020,kaur2025optimizedquantumcircuitpartitioning}, partitioning the high-level Pauli strings allows us to obtain more dense partitions because many circuit complexities are abstracted away.

\begin{algorithm}[t]
\caption{Greedy Partitioning Algorithm}
\label{greedypartitionalgorithm}
\begin{algorithmic}
\Require List of Hamiltonian terms $Hamiltonian\_terms$
\Ensure Partitions of Pauli operators acting on at most 3 qubits

\State \textbf{Sort} $Hamiltonian\_terms$ by their highest qubit index then by term weight
\State $partitions \gets []$
\For{each $term$ in $sorted\_terms$}
    \State $placed \gets False$
    \For{each $partition$ in $partitions$}
        \If{combined qubits of $term$ and $partitions$ contain at most 3 qubits}
            \State \textbf{append} $term$ to $partition$
            \State $placed \gets True$
            \State \textbf{break}
        \EndIf
    \EndFor
    \If{not $placed$}
        \State \textbf{append} $[term]$ as a new partition to $partitions$
    \EndIf
\EndFor
\State \Return $partitions$
\end{algorithmic}
\end{algorithm}

Our Hamiltonian term partitioning algorithm is shown in Algorithm~\ref{greedypartitionalgorithm} and we explain it using the example in Table ~\ref{fig: greedypatition}.
In this table, the input is the terms of a 4 qubit spin Hamiltonian where each term is weight 1 or weight 2. 
After receiving the input, the terms are ordered by the largest qubit index acted upon in the term. 
The terms are then sorted by weight when two terms have an identical max index to define the final ordering. 

For example, consider Pauli string $X_1X_2X_3$ and $Z_3$. 
The highest qubit index of both terms is shared, and therefore what would decide the final ordering is the weight of the terms (i.e., $Z_3 \leq X_1X_2X_3$).
In this sorted order, locally overlapping or anti-commuting terms that should be partitioned together effectively appear near each other, while high-weight or irrelevant terms end up at the tail of the array. 
In Table~\ref{fig: greedypatition} we see that $Z_1$ and $X_1X_2$ are non-commuting and naturally align close to each other because non-commutation is determined strongly by shared indices. 
Due to many Hamiltonians being local in nature, sorting by qubit indices tends to put large portions of non-commuting terms very close to each other in the array. 

The partitioning phase uses a greedy algorithm which adds terms to the first partition it sees available. If no half constructed partition is available, a new one is created. 
In our example, $Z_1$ will invoke a partition creation, $X_1X_2$ and $X_3$ will then be added to the same partition. 
At this point the group is full, so when $X_3X_4$ is selected next going from left to right, a new group will be created to avoid having more than 3 unique indices in one group. 
The resulting partitions tend to be dense enough to allow meaningful circuit optimizations while also maintaining simplicity. 

\subsection{Trotter Step Reordering and Randomization
\label{sec:Trotterreorder}
}

\begin{figure}
    \centering
    \includegraphics[width=\linewidth, height=4cm]{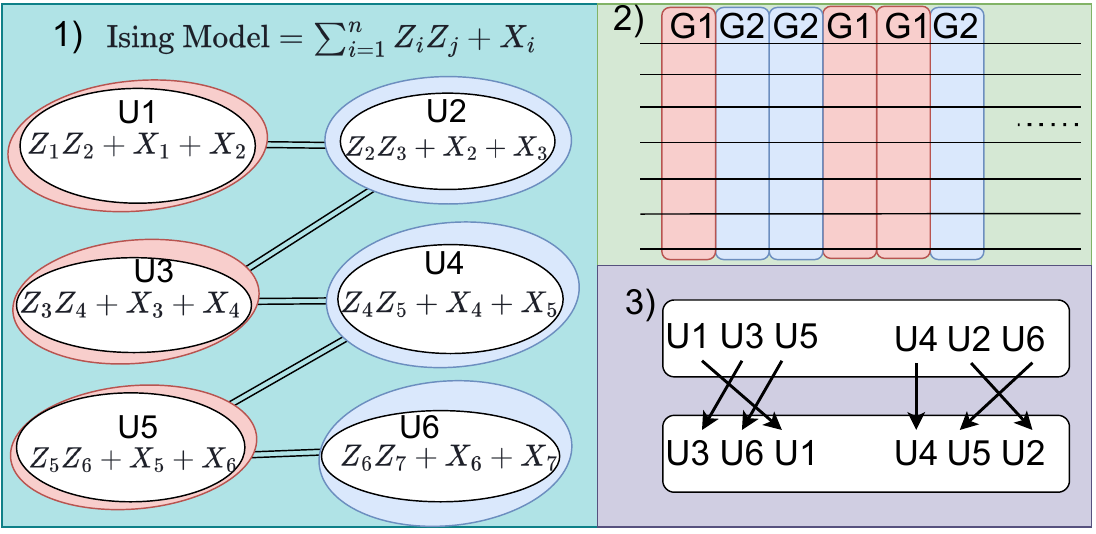}
    \caption{1) \textbf{Create Groups:} A conflict graph is constructed showing commutation relations of Hamiltonian terms. A vertex indicates a unitary of the Trotter step. An edge indicates that two unitaries do not commute. Independent sets are created about the graph which are used to group unitaries with other pairwise commuting unitaries. 2) \textbf{Order Full Groups:} The groups created are ordered in the Trotter step for cancellation with other groups. The two largest groups are placed on edges of the Trotter step. At the neighboring Trotter steps, the groups placed at the edges swap places such that identical groups are neighboring each other. Unitaries are then merged via commutation equivalences. 3) \textbf{Shuffling Group Term Order:} The order of terms within each group is shuffled to invoke stochastic noise over coherent noise. }
    \label{fig:Trotterreorder}
\end{figure}

In the second stage of our optimization pipeline, we reorder and randomize our partially Trotterized unitaries (see Figure \ref{fig:Trotterreorder}).
Here, we use a simple input case, the 1D Ising model. 
The input consists of a set of partially Trotterized unitaries of the form $e^{i(\sum H_i)t}$, which together form a single Trotter step. The Pauli strings contained in each unitary are also listed.
In Step 1, we construct a conflict graph that represents the commutation relationships between the Trotter step unitaries. 
These unitaries are generated as outputs from the previous algorithm described in section 4.1.
Independent sets, corresponding to mutually commuting unitaries, are then extracted from this graph to form commuting groups. 
The two independent groups are denoted as G1, and G2 respectively, in Figure \ref{fig:Trotterreorder}. 
Extracting independent sets is done in a greedy fashion according to Figure. \ref{fig:Trotterreorder}.
After identifying independent sets, Step 2 shows the ordering of groups within 1 Trotter step. 
Groups are ordered such that with neighboring Trotter steps, identical groups are neighboring each other and can be trivially merged into fewer unitaries; this is beneficial for the final output.. 
For example, imagine $e^{iH_it}$. 
Due to all of the terms mutually commuting, the identical unitaries can be reordered such that $e^{iH_i t} e^{iH_i t} \to e^{i2H_it}$  which reduces the unitary count from the perspective of mapping unitaries to gates. 
Step 3 we mitigate coherent noise by shuffling the order of unitaries in each group.
Notice that the ordering is not shuffled between groups, and that all unitaries stay within their assigned group from Step 1. 
This approach effectively reduces the overall circuit depth and gate complexity, optimizing the quantum circuit compilation without incurring additional approximation errors.

Here we describe how to obtain the groups found in Step 2 of Figure \ref{fig:Trotterreorder}. 
The greedy independent set algorithm, described in Algorithm \ref{Trotterstepreordering} starts with the conflict graph as input. 
Starting with a vertex, for example the vertex with the lowest index, add all vertices not sharing an edge with the target vertex to our group. 
Second, we need to remove all vertices in our newly formed group from the conflict graph so that these vertices are not repeated in newer groups. 
The process is then iterated again 
 to get the second largest maximally independent set of the graph. 
The conclusion of this algorithm outputs two sets which are to be merged with their identities on the boundaries of Trotter steps, as seen in Figure \ref{fig:Trotterreorder}, Step 2. 

\begin{algorithm}[!]
\caption{Trotter Step Reordering and Randomization}
\label{Trotterstepreordering}
\begin{algorithmic}
\vspace{0.5em}

\Function{BuildConflictGraph}{$H$}
    \State Initialize graph $G = (V, E)$ where each node $v_i \in V$ corresponds to a term in $H$
    \For{each pair of terms $(t_i, t_j)$ in $H$}
        \If{$[t_i, t_j] \neq 0$ \textbf{(they do not commute)}}
            \State Add edge $(v_i, v_j)$ to $G$
        \EndIf
    \EndFor
    \State \Return $G$
\EndFunction
\vspace{0.5em}

\Function{GreedyCommutingGroups}{$G$}
    \State $groups \gets []$
    \While{$G$ is not empty}
        \State $I \gets \text{GreedyMaxIndependentSet}(G)$ 
        \State \textbf{append} $I$ to $groups$
        \State Remove nodes in $I$ (and their edges) from $G$
    \EndWhile
    \State \Return $groups$
\EndFunction
\vspace{0.5em}

\Function{ReorderTrotterSteps}{$\{H_1, \dots, H_n\}$}
    \For{each Trotter step $H_k$}
        \State $G_k \gets \Call{BuildConflictGraph}{H_k}$
        \State $groups_k \gets \Call{GreedyCommutingGroups}{G_k}$
        \State \textbf{randomize} the ordering \textbf{within} each group in $groups_k$
        \State \textbf{concatenate} commuting groups contiguously 
    \EndFor
    \State \textbf{reorder} consecutive Trotter steps
    \State \textbf{merge} commuting operators across adjacent steps where possible:
        \If{$[A, B] = 0$ for $A$ in step $k$, $B$ in step $k{+}1$}
        \EndIf
    \State \Return \{modified Trotter steps\}
\EndFunction

\end{algorithmic}
\end{algorithm}

\subsection{Unitary Decomposition for Grouped Hamiltonian Terms}


After we group the Hamiltonian terms and order them, the final step is to decompose these grouped terms into basic gates.
As discussed in section~\ref{sec:opporutnities}, the key to successfully leveraging the benefit from partitioned Hamiltonian terms is being able to efficiently decompose the exponential of the partitions into basic gates.
An MCTS is an algorithm designed to handle sequential decision problems where there is little information about the environment, which is exactly the problem of circuit synthesis for general combinations of Hamiltonian terms. 
With a good balance of exploring new solutions and exploiting known working solutions, performance can be better than greedy heuristics and have more flexibility than dynamic programming-based approaches. 

\begin{figure}
    \centering
    \includegraphics[width=0.9\linewidth]{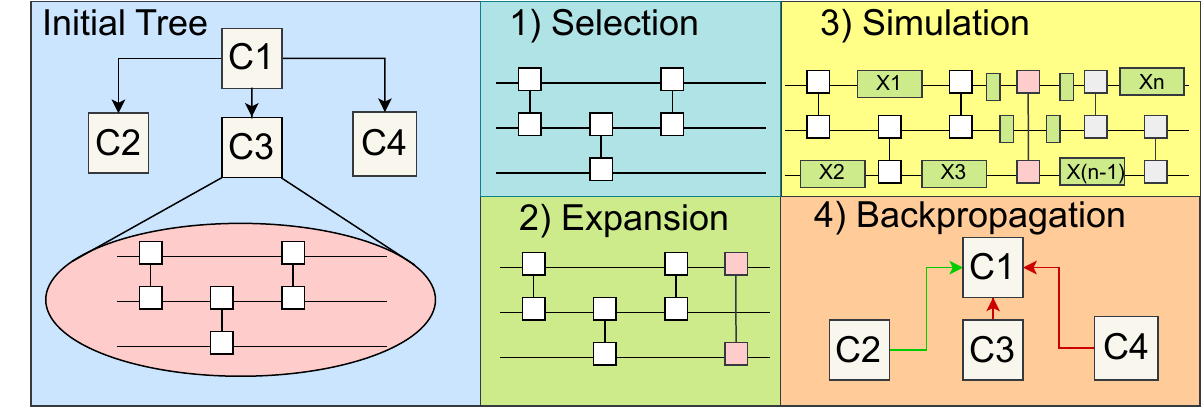}
    \caption{\textbf{Unitary decomposition method} 1) \textbf{Selection:} Select a node in the search tree which represents a partially synthesized circuit which has unexplored child actions. 2) \textbf{Expansion:} Select a CNOT gate among choices from the gateset to append to the circuit. 3) \textbf{Simulation:} Starting from the newly expanded state, append CNOTs until we reach a terminal circuit length. After, interleave a fixed number of single qubit gates at random in between the CNOT gates. Optimize parameters with the Gauss-Newton method. 4)\textbf{ Backpropagation:} Update values of nodes in the tree based on the result of the simulation stage to identify if the newly explored state was valuable.}
    \label{fig:mctsimp}
\end{figure}

An example of how MCTS elements fit into our framework is shown in Figure \ref{fig:mctsimp}.
Referring to the initial tree in the example, each tree node state is a circuit of strictly CNOTs. 
Actions the algorithm can take are defined as CNOT gates which can be appended to a partially synthesized circuit expressed by a node state.   
The MCTS algorithm starts with the \textbf{selection} process. 
The goal of selection is to find a promising node of the tree data-structure for which actions taken from that node state have not been explored yet.
During our selection process, we traverse the tree using a policy until we reach a node with unexplored actions.
The circuit shown in blue is the partially synthesized circuit for which the node selected represents. 
In the \textbf{expansion} step, an unexplored action is explored which leads to a new node being appended to the tree as a child to our selected node. 
The difference now is a CNOT gate has been appended to our selected node state, creating a new state that has no known value yet. 
In the \textbf{simulation} step random CNOT gates are then appended to the circuit. 
Following the appending of random CNOT gates up to a fixed circuit length, single qubit gates are then interleaved between all CNOTs. 
The result is the circuit diagram shown in the simulation step of Figure. \ref{fig:mctsimp}. 
After generating a fully synthesized circuit, parameters of single qubit gates are solved for such that the values minimize the error between the synthesized circuit and the target unitary. 
The value of the state is then determined by the amount of CNOT gates and the error of the approximation. 
At the end of our algorithm, the fully synthesized circuit with the largest value is returned.
\textbf{Backpropagation} is the final stage of the algorithm where the value of each state is updated based on the results of the simulation stage. 
In the example, three partial circuits were evaluated and the values of the results are passed from the leaf nodes to the root node. This allows the algorithm to learn and make better decisions on future iterations. 

To select a node, a key tradeoff in the field of reinforcement learning is the balance of exploiting known solutions and exploration of new solutions that may lead to better results. 
The selection of a node to explore is determined by a policy. 
A policy in general context is how the algorithm decides which actions to take.
For our policy, the input would be the value of nodes to traverse to and the number of times the nodes have been explored.
The output is a decision of which action to take leading to the state deemed most promising by the policy. 
In Monte Carlo search tree, a common policy for this purpose is the canonical UCT policy~\cite{sutton2018reinforcement}.




Why focusing on CNOT skeletons.
When designing the MCTS algorithm, we must consider the concept of coverage. 
Coverage is a quantity that says how much of the search space can be covered by the RL algorithm.
Quantitatively, coverage can be defined as follows: 
\begin{equation}
\mathrm{Coverage} \approx \frac{N_{\text{visited}}}{b^{D_{\max} + 1}}
\end{equation}
Here, $N$ is the number of states visited, $b$ is the number of choices available at each state (the branching factor), and D is the maximum possible length of a sequence (a circuit in our case). 
In order to make the environment more tractable, our depth and branching factor should be as small as possible which will give more coverage of the search space to find an optimal solution. 
In the straight forward framework of circuit synthesis, our branching factor (due to choice of angles) and the depth of a circuit can be impractical for light weight RL agents due to continuous action spaces and very long depths. 

A key insight to our algorithm design is that we only consider CNOT gates when defining states of the partially synthesized circuit. 
The motivation was out of necessity to condense the search space of synthesizing a circuit where the search space is defined by all permutations of a universal target gate set. 
The intuition is that the entanglement structure is the most difficult characteristic to solve in circuit synthesis and that single qubit gates that are continuously parameterized can lead to a smooth landscape for optimization via differentiable methods.
For our approach, once an entanglement structure is determined, the circuit is overparameterized with many single qubit gates injected at all circuit layers. 
Overparameterization is important because it leads to a smoother cost landscape compared to a function with fewer parameters. 
Using the Gauss-Newton method, we minimize the L2 norm, our cost function, of the difference matrix between the target and approximation circuit. 
After getting an optimized solution, all strings of single qubit gates can be rewritten as one single qubit gate making the circuit optimal for quantum hardware. 
For our implementation, the Qiskit transpiler at level 3 optimization is used to convert our overparameterized circuit into an optimal circuit expressed in the (u3,cx) gateset. 

Value of our simulated solution is calculated as a function of accuracy and gatecount (Eq. 2). 
The function is non-continuous and depends on the accuracy of the circuit being above or below a threshold error, which we have set to $10^{-8}$. 
If the error of the approximation after simulation is below this threshold, value is determined strictly by the negative of CNOT gate count. 
However, if the error of the approximation is above the threshold, value is determined strictly as negative error. 
For example, if the circuit in Simulation of Figure. \ref{fig:mctsimp} had an error of below $10^{-8}$, then the value would be -6. However, if the error was above the threshold, the value would be $-\epsilon$. 
\begin{align}
\mathcal{E}(x) &= \text{argmin}_{\theta} || \prod_{i=1}^{n} x_i(\theta_i) - U ||_2 \\
R(x) &= \begin{cases}
    -\text{\#cnot}, &\mathcal{E}(x) < \epsilon \\
    -\mathcal{E}(x), &\text{otherwise} 
\end{cases}
\label{eqs:costfunction}
\end{align}
The intuition is that there will be important information, referred to as a signal, given even in the event of failed simulations to tell the algorithm where more and less accurate solutions are occurring. 
After finding solutions over a threshold, accuracy offers diminishing returns and gatecount becomes a larger priority. Backpropagation is then simply performed by updating all $Q_i$ from the UCT policy for each node that has been traversed in the selection phase. If a winner is found, in practice many are found at once, then the best circuit is returned immediately. 

\section{Evaluation
\label{sec:Evaluation}}

\begin{table}[t]
    \centering
        \caption{Benchmark information with grid sites and final qubit counts (for Fermi–Hubbard, qubit\# = 2 $\times$ grid sites)}
    \resizebox{\columnwidth}{!}{\begin{tabular}{|l|l|l|l|}
    \hline
    \textbf{Benchmark} & \textbf{Topology} & \textbf{ Size} & \textbf{Qubits} \\ \hline
    \multirow{3}{*}{\textbf{Fermi-Hubbard (FH)}} 
        & Triangular Grid & $2 \times 2$ & 8-128 \\ \cline{2-4}
        & Square Grid     & $2 \times 2$ & 8-128 \\ \cline{2-4}
        & 1D Grid         & $5 \times 1$ & 10-144 \\ \hline
    \multirow{3}{*}{\textbf{Heisenberg (HB)}} 
        & Triangular Grid   & $5 \times 2$ & 10-144 \\ \cline{2-4}
        & Rectangular Grid  & $5 \times 2$ & 10-144 \\ \cline{2-4}
        & 1D Grid           & $10 \times 1$ & 10-144 \\ \hline
    \multirow{3}{*}{\textbf{Ising (IS)}} 
        & Triangular Grid   & $5 \times 2$ & 10-144 \\ \cline{2-4}
        & Rectangular Grid  & $5 \times 2$ & 10-144 \\ \cline{2-4}
        & 1D Grid           & $10 \times 1$ & 10-144 \\ \hline
    \textbf{LiH Molecule (LiH)} 
        & Molecular         & N/A &  10\\ \hline
    \textbf{HF Molecule (HF)} 
        & Molecular         & N/A & 10 \\ \hline
    \textbf{PD-1 Protein (PD1)} 
        & Molecular         & N/A & 28-222 \\ \hline
    \end{tabular}}

    \label{tab:systeminfo}
\end{table}

\begin{figure*}[htbp]
    \centering
    \includegraphics[width=1\linewidth, height=11cm]{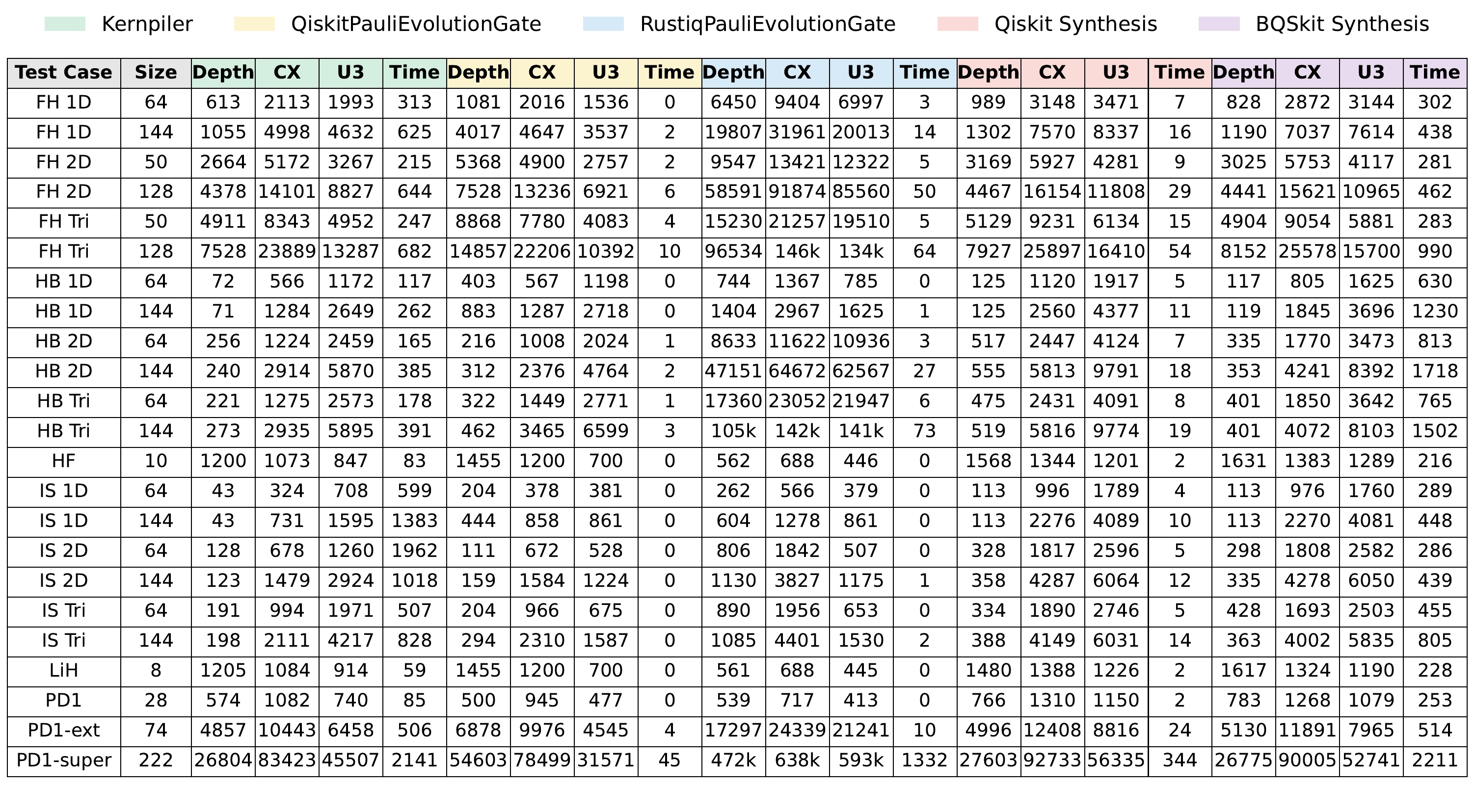}
    \caption{Absolute data comparisons with no error reduction considered. Hamiltonians are compiled to a fixed number of Trotter steps. 
    }
    \label{fig:runtimescalability}
\end{figure*}

\textbf{Experimental Configuration:} To evaluate our work, we measured performance using metrics with and without error scaling accounted for. 
For error quantification, we compare the approximate unitary with the theoretical perfect unitary using the L2 norm of simulation Hamiltonians that involve between 8 and 10 qubits. 
The L2 norm has been commonly used to quantify error of approximations in quantum algorithms \cite{dawson2005solovaykitaevalgorithm,Childs_2021} and therefore that is our metric of accuracy here.
Our target is to compile results to greater than $99.5$\% accuracy.
We also perform Trotterization comparisons to evaluate error reductions in both near-term and long-term applications. For the second order Trotterization, we use a time simulation with t = 1 in dimensionless units, and our experience shows that for significantly longer simulations, the second order method performs markedly better for most general tasks compared to the first order Trotterization. 
In contrast, for the first order Trotterization, we consider short time simulations by scaling all Hamiltonian coefficients by t=0.1 which is appropriate for near term applications to observe short time dynamics of quantum systems. 
It is important to note that Qiskit's PauliEvolutionGate currently defaults to first order Trotterization; therefore, we recommend viewing the corresponding chart for a more accurate state-of-the-art comparison and review the second order for future more general use of quantum computers for quantum simulation. Scalability is assessed by measuring runtime and gate count using 28-220 qubit Hamiltonians. 
For the larger Hamiltonians, the L2 norm cannot be measured however we expect the same reduction in error at larger sizes because the weights of terms do not increase with system size for most Hamiltonians.
We discuss the scalability in Section \ref{sec:errorreduction}

\textbf{Software and Hardware Setup:} Our implementation is carried out using PyTorch version 2.5.1+ CUDA 12.1, and we compare our results against Qiskit's stable version 1.3.2, which features state-of-the-art Hamiltonian compilation methods inspired by the works of Rustiq \cite{debrugiere2024fastershortersynthesishamiltonian} and Paulihedral \cite{li2021paulihedralgeneralizedblockwisecompiler}. The hardware setup includes an A100 GPU with 80GB of RAM for implementing the Monte Carlo search tree, alongside an AMD EPYC 9654P 96-Core Processor for the overall implementation. 
Furthermore, for evaluation of our Monte Carlo Unitary Synthesis, we compare against some existing the unitary synthesis toolkits in Qiskit stable version 1.3.2 and BQSkit \cite{bqskit} version 1.2.0.

For circuit generation, we create Qiskit circuits for all algorithms, including our proposed method, the paulievolutiongate, and the paulievolutiongateRustiq. 
In the case of first-order Trotterization, we employ the Lie-Trotter formula, modifying only the number of steps from the default configuration. 
For second-order Trotterization, we use the Trotter-Suzuki formula with the same adjustment in the steps argument. After circuit generation, we optimize the circuits at level 3 optimization in Qiskit's transpiler using the u3 and CNOT basis with all-to-all connectivity. 
The optimized circuit is then converted into a numerical format to calculate the L2 norm of the difference matrix, and by squaring this norm, we estimate the order of magnitude on state fidelity.

\textbf{Benchmarks:} To ensure a comprehensive evaluation, we select a wide range of popular Hamiltonians that vary in topology, geometry, terms, and correlation structures (see Table \ref{tab:systeminfo}).
For nearest neighbor models, we include the Ising and Heisenberg models, which demonstrate varying site densities (the number of Hamiltonian terms per site). Additionally, we consider non-local models, such as the Fermi-Hubbard model and molecular Hamiltonians, where variations in correlation and dimension help expose the strengths and weaknesses of the different compiler methods.
All fermionic models have been mapped to qubits using the Bravyi-Kitaev mapping \cite{seeley2012bk}

\subsection{Comparison with Baseline without Considering Error Reduction
\label{subsec:absolutecomparision}}

Firstly, we will discuss the absolute data comparisons in terms of gate count and circuit depth when using various compilation techniques to quantify compilation efficiency per Trotter step rather than the accuracy of the compilation.
That is, the compilation error reduction is not included and will be evaluated later in the next section.
Therefore, different from other experiments, our Hamiltonian sizes are of range 28-220 qubits, and the L2 norm was not considered. 
For competitive benchmarks, we compare against two state-of-the-art unitary synthesis techniques: BQSkit and Qiskit's unitary synthesis, and two state-of-the-art fully integrated Hamiltonian compilation schemes: Rustiq and Paulihedral. 
For the unitary synthesis, we replaced the Monte Carlo Search Tree algorithm from Section 4 with BQSkit and Qiskit to compare ablated performance to our full Kernpiler approach. 
For metrics compared, each Hamiltonian was compiled to 3 Trotter steps, for which we recorded the quantity of U3 and CNOTs within the end-to-end circuit. Other metrics besides gatecount measured were the system size, compilation time (seconds), and total depth of the circuit. 
This data is summarized in Figure \ref{fig:runtimescalability}.

\begin{figure}
    \centering
    \includegraphics[width=\linewidth, height=6cm]
    {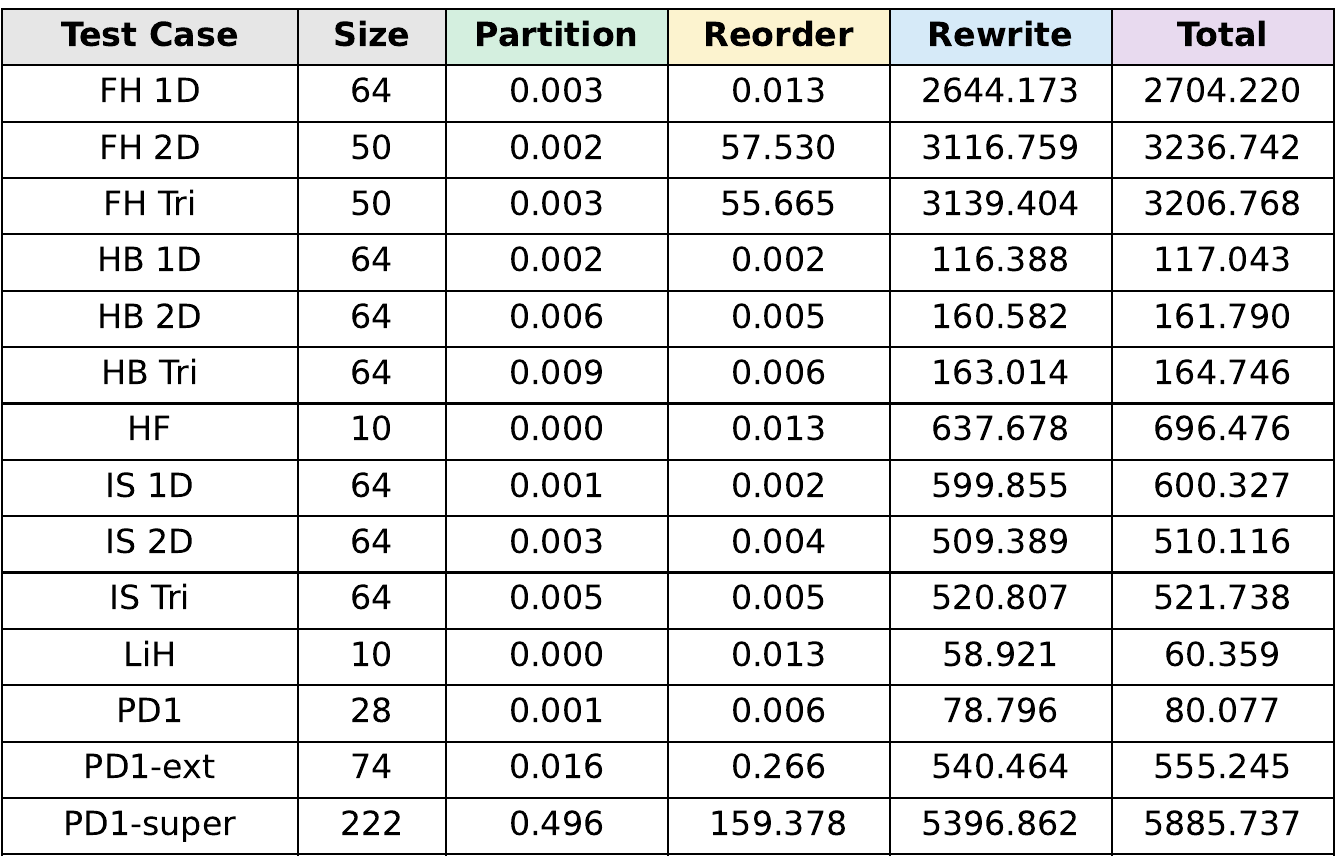}
    \caption{ Runtime (in seconds) for all passes of Kernpiler when compiling large benchmarks}
    \label{fig:runtimedetailed}
\end{figure}

We now summarize our interpretation of Figure \ref{fig:runtimescalability}. 
Our MCTS-based approach achieves the smallest gate counts among the other unitary decomposition techniques without having the longest runtimes, akin to BQSkit, for which we cached the compiler to save on time. 
More specifically, compared to the ablated MCTS pass, our results achieve a peak(average) reduction of U3 gate counts by 60\% (33\%), CNOTs by 67\% (34\%), and depth by 60\%(32\%) compared to the next best performer for each benchmark. 
The reason BQSkit and our MCTS-based method have longer compilation times compared to other existing techniques is due to the fact that BQSkit and our methods are search-based, where the algorithms will try different decompositions in a large search space to try and find an optimal decomposition.
Other approaches are based on constructive heuristic algorithms.
For a breakdown of runtime per pass, see Figure~\ref{fig:runtimedetailed}. 

Our approach outperforms BQSkit in runtime by 
leveraging implicit bias in the search space. 
By having high-level information given to the MCTS, our compiler can perform a more reasonable search of available options and can also set obtainable optimization goals (such as searching for a 10-gate solution vs a 5-gate solution). 
Additionally, BQSkit and Qiskit both have the issue that they are performing unitary synthesis at the lowest level of representation and cannot assume a high-level structure of their input, allowing for goal setting, which reduces the search space. 

An exciting result is that our technique has advantages even when excluding the Trotter steps saved. 
Furthermore, our technique still achieves a peak(average) CNOT reduction of 14\% (0\%) and a depth reduction of 60\%(14\%) compared to the next-best-performing technique for each benchmark. 
The Kernpiler design also manages to close absolute gate performance gaps as we increase the size and complexity of our Hamiltonian. 
For example, our most complex benchmark (PD-1 protein fragments) has a large performance gap for small sizes of 28 qubits and progressively closes the gap in gate count and achieves the best depth metric in the largest fragment of 222 qubits.
Further, we point out a weakness in the U3 count, without considering Trotter step reduction, U3 gate count increases by 50\% on average compared to the next best performer, but is handled by saved Trotter steps which we show later. 

A reason for our results in depth is due to the design. 
By not having optimizations primarily focused on reordering the Trotter step, which invokes a tradeoff between depth and gate count, our stack can optimize along the axis of depth and gate count concurrently. 
Therefore, while partial Trotterization and MCTS optimize gate count, reordering can be used to optimize depth, which avoids the original tradeoff.

We now discuss a limitation of our design regarding the lack of optimization towards single qubit gates. 
There are two systemic reasons for this. 
Firstly, our MCTS is optimizing over CNOT count rather than total circuit complexity, as illustrated by the cost function of Equation \ref{eqs:costfunction}. 
Secondly, we intentionally over parameterize the circuits with single qubit gates to gain convergence. 
This over parameterization is then optimized via existing compilation techniques. 
The combination of injecting complexity into the circuit and not designing the search space to express single qubit gates is contributing to the lack of performance on this metrics. 
Such limitation can be mitigated by more complex search heuristics which include U3 gates as an optimization target.

\begin{figure*}[htbp]
  \centering
  \includegraphics[
    width=.8\linewidth,
    height=7cm
  ]{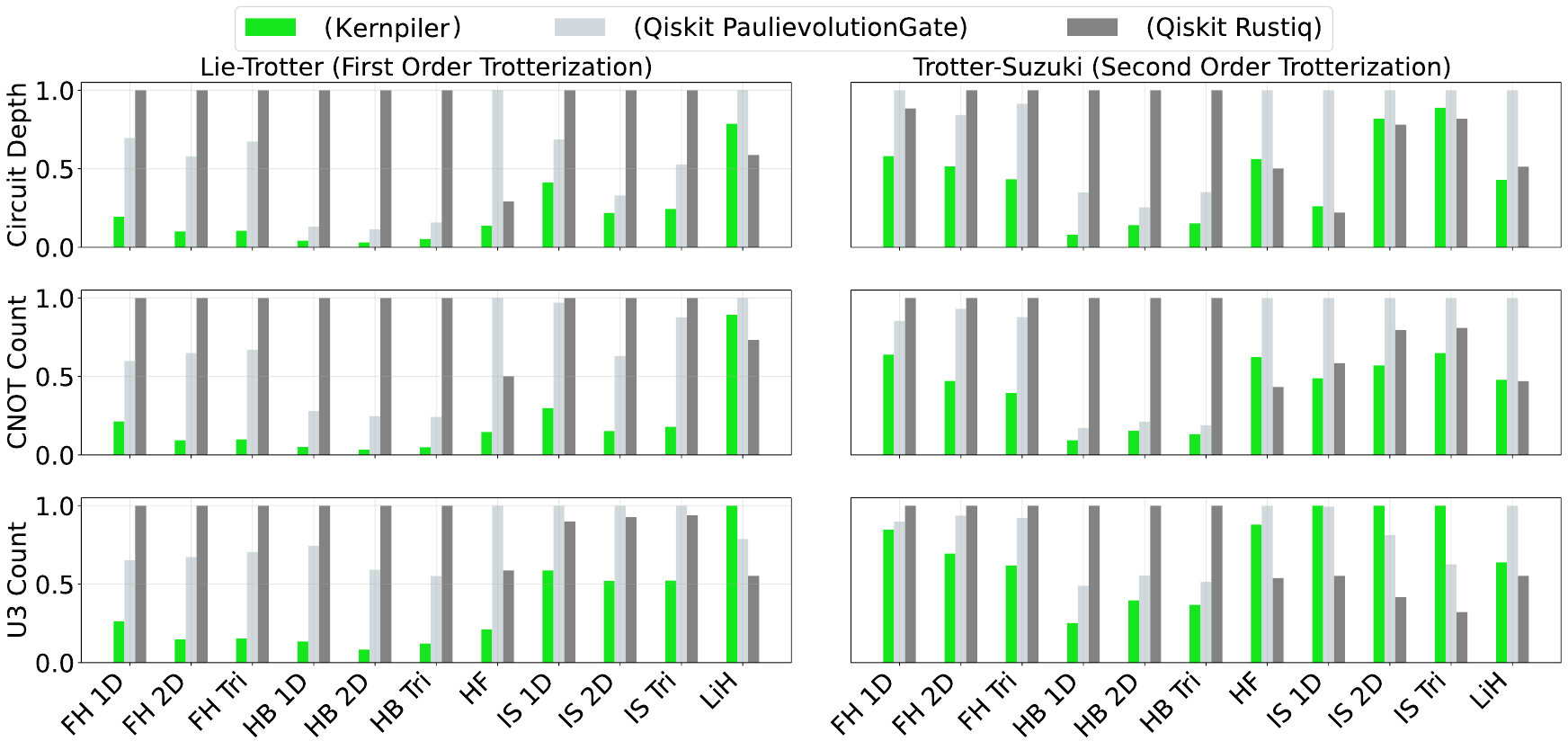}
  \caption{Depth, CNOT and U3 count comparison when compiling to less than 1\% approximation error on a range of time evolution unitaries.}
  \label{fig:datascaling}
\end{figure*}

\subsection{Comparison with Error Measurements}

We now discuss the comparison of our compilation software against competitive benchmarks focused on Trotterization optimization when considering error savings. Figure \ref{fig:datascaling} presents the results for first-order Trotterization (Lie--Trotter) and second-order Trotterization (Trotter--Suzuki). 
The graphs are normalized to display percentage reductions from the maximum gate count(depth) observed. 
Overall, the data reveals a higher reduction for the first-order Trotterization compared to the second order, which still achieves about a three-fold reduction in the best-case scenarios for gate count and up to a 10x reduction in depth.

Two primary factors account for the difference between first and second-order improvements. 
First, the constant factor in our commutation relation is reduced by a square root for the first-order Trotterization. 
Specifically, while the first-order Trotterization scales as \(\Delta t^2/N\), the second-order scales as \(\Delta t^3/N^2\) where $N$ is the number of Trotter steps. 
Consequently, a constant reduction factor in the numerator, which is what partial Trotterization enables, will be diminished by an \(N^2\) step scaling in the second order case, whereas the first order requires a linear number of steps, as compared to a square root number of steps, to reach the same level of optimization. 
Second, for bipartite Hamiltonians—those whose conflict graphs from Section 4.2 are bipartite—the two commutator groups span a large portion of the Trotter step. 
Because the order of commuting groups is reversed in these cases, an almost \(\Delta t^3\) scaling can be observed. This behavior, evident across a wide range of benchmarks, is attributed partly to system size and partly to the high degree of commutativity. 
These effects also explain why, in the second-order data, the reduction does not reach the square root improvement observed in the first-order Trotterization, as the competition scales more appropriately with our method. Additional observations include the performance differences among the various compilers. 

Qiskit's PauliEvolutionGate tends to perform best on very regular, low connectivity, low weight Hamiltonians, while Rustiq performs , by design, optimally on molecular/electronic structures with non-trivial connectivities and terms.
The largest gap in performance is observed in cases with non-trivial yet regularized connectivities, such as the triangular lattice and electronic Hamiltonians with long-range correlations over a symmetric lattice. Additionally, our compiler tends to perform very well on Hamiltonians that are denser in terms per site (i.e the Heisenberg models vs the Ising models). 
This outcome can be attributed to the nature of our optimizations; relatively local connectivity—even in the presence of non-trivial topologies—allows our grouping algorithm to identify large commuting sets, and our rewrite procedures, being independent of other Hamiltonian terms, are less affected by unpredictable correlations. 
Notably, Rustiq appears to underperform on most non-electronic Hamiltonians. In contrast, PauliEvolutionGate serves well as a general spin Hamiltonian compiler, excelling on symmetric local connectivity but struggling with irregular patterns, as evidenced by its performance on electronic structure Hamiltonians and the atypical topologies found in local/power law Hamiltonians.

For the Ising models, an interesting discrepancy is observed: while the CNOT gate count is extremely low, the U3 count is significantly higher. This is because our rewrite system does not employ a CNOT tree or chain for decomposition. 
As a result, more U3 unitaries appear in odd or sandwiched locations, whereas a CNOT tree decomposition would eliminate the need for basis changes and require only a single Z gate, thereby intrinsically reducing the U3 count.

\section{Error Reduction Theoretical and Experimental Data
\label{sec:errorreduction}} 

Here we offer a theoretical explanation for the error reductions observed, alongside an understanding of how this concept scales to larger rewrite radii and lattice size. Theoretical error reduction fundamentally arises through commutator cancellations. To illustrate this, we start from the standard derivation of Trotterization, where the error terms can be expressed as a sum of commutator norms:
\begin{equation}
\text{Error} = \sum_{i<j} \frac{|[H_i,H_j]|}{2} \Delta t^2 + \mathcal{O}(\Delta t^3).
\label{eqs:commutation_error_theory}
\end{equation}

By partitioning Hamiltonian terms, we instead consider commutators between entire groups rather than individual terms, leading to:
\begin{align}
\text{Error}{~\text{partitioned}} &= \sum_{A<B} \frac{|[H_A,H_B]|}{2} \Delta t^2 + \mathcal{O}(\Delta t^3) 
\end{align}
where each group $H_A$ is composed of individual Hamiltonian terms maximized for non-commutativity. Importantly, the commutator between partitions $[H_A,H_B]$ is simply the aggregation of all individual commutators $[H_i,H_j]$ where  $H_i \in H_A$ and $H_j \in H_B$. 
Thus, the partitioned error (Eq. 8) explicitly represents the original error minus the intra-group commutator contributions that vanish due to partially Trotterized unitaries. 
This leads to a final reduced error of Trotterization to: 

\begin{equation}
\text{Error}{ \text{ reduced}} = \text{Error} - \text{Error grouped},
\end{equation}

quantifying the precise error savings achieved through term partitioning and highlighting the scalability of this methodology.
As the partition size increases, the number of intra-partition commutators grows combinatorially, scaling roughly as  $n_A^2$ for a partition of size $n_A$. Consequently, error reduction becomes significantly more pronounced as larger partitions are formed, since more commutator terms vanish. 
Thus, increasing the rewrite radius directly enhances error reduction, emphasizing the scalability and efficiency of this partial Trotterization approach in practical quantum simulations.

\begin{figure}
    \centering
    \includegraphics[width=0.95\linewidth, height=7cm]{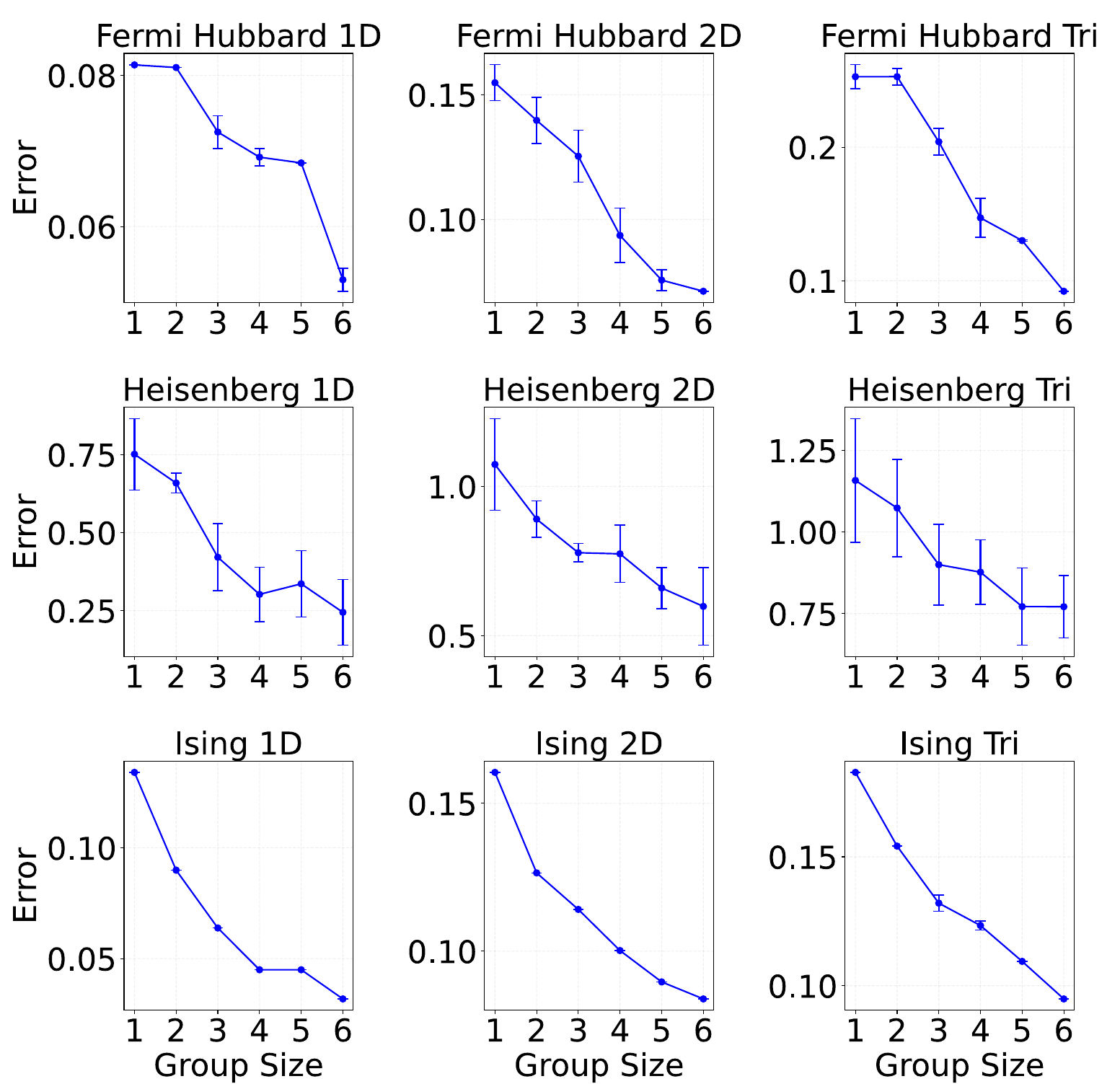}
    \caption{Increasing the number of qubits per unitary to decompose directly reduces the error.}
    \label{fig:scalingdata}
\end{figure}

We investigated this empirically with first order Trotterization of Hamiltonians decomposed using 10 Trotter steps with no special optimizations.
The only change over decompositions is the amount of partial Trotterization performed. 
In Figure \ref{fig:scalingdata}, we show scaling of the compiler error versus group decomposition size (number of qubits) across 3 different models with 3 different geometries. We performed 5 runs per data point. The remarkable find is that the approximation error decreases drastically as a function of group size; this highlights a remarkable benefit of the partial Trotterization schema.

The Ising models possess the monotonic trends which are likely an artifact of the simple distribution of Hamiltonian terms that allows for easily converging on the best partitions. 
For the other models we see more significant effects from noise. This originates from the partitioning of Hamiltonian terms. 
As the entanglement structure becomes increasingly non-trivial, the partitioning algorithm encounters greater difficulty converging to the optimal partitions causing more noise in the commutation error observed. 

\textbf{Scalability Discussion}
Now we discuss the scalability of our technique to larger quantum lattices and Pauli weight terms. 
Modeling the exact error is intractable as we scale qubit size, however Trotter error is directly proportional to the amount of non-commuting pairs of terms which define the Hamiltonian (see Equation \ref{eqs:commutation_error_theory}).

\begin{figure}
  \centering
  \includegraphics[
    width=.95\linewidth,
    height=8.2cm,
  ]{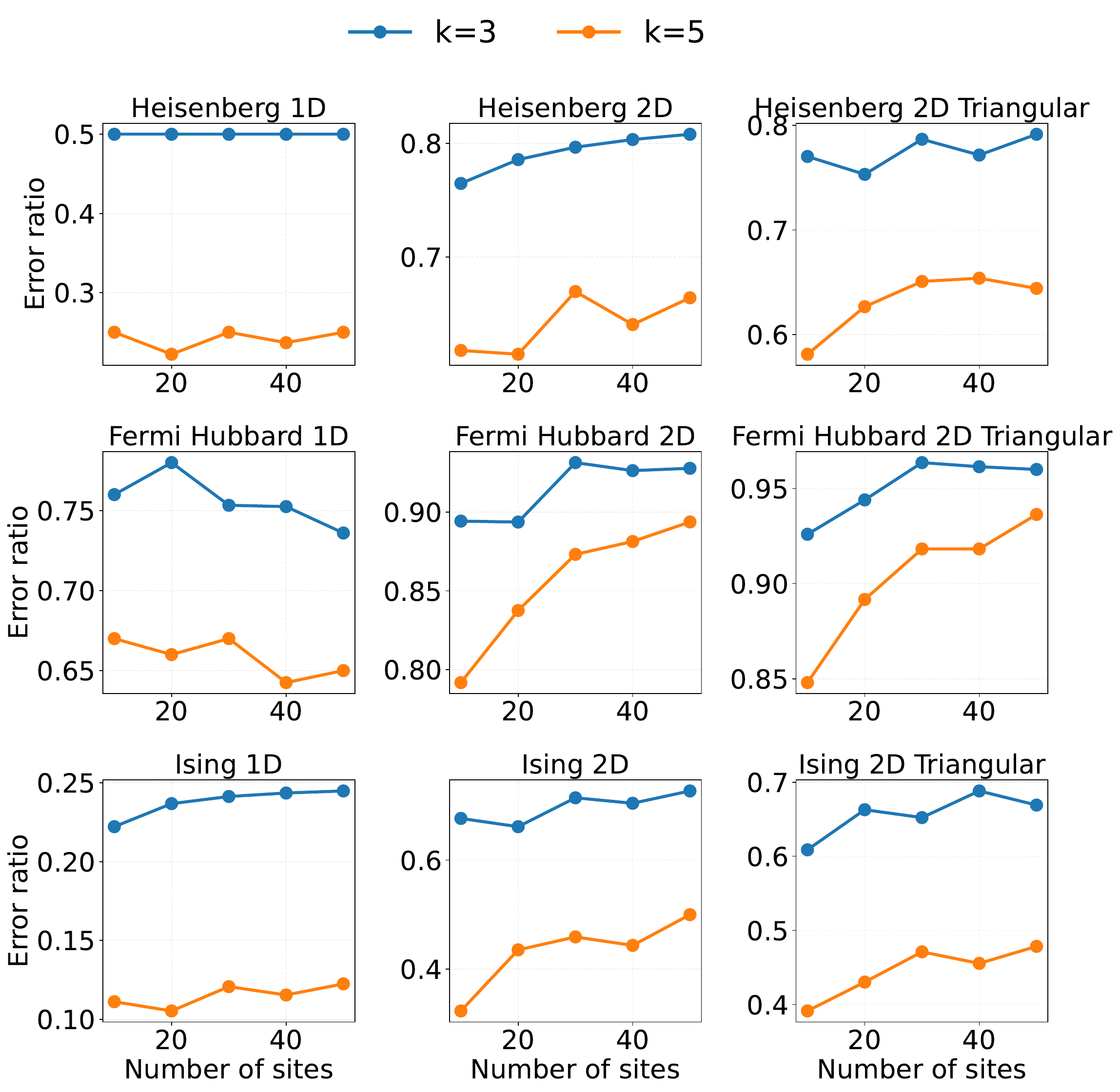}
  \caption{Ratio quantifying percentage reduction of non-commuting pairs of Hamiltonian terms between partitioned and unpartitioned Hamiltonians over increasing quantum lattice sizes.}
  \label{fig:scalability_proxy}
\end{figure}

Figure~\ref{fig:scalability_proxy} shows the amount of non-commutivity as we increase qubit size.
In this experiment, we graph the ratio of non-commuting pairs between a partitioned and unpartitioned Hamiltonian. 
More specifically, we graph $\text{\#non commuting pairs partitioned} \over \text{\# non commuting pairs}$.
Partition sizes used were $n=3$ and $n=5$. 
This ratio is measured over system sizes from 10 to 50, extending out from our benchmarks measured in Figure~\ref{fig:datascaling}.
At each qubit array size, we measure the ratio of non-commuting pairs, seen on the Y axis.
The expected behavior is that for k-local Hamiltonians, the non-commutation ratio should not significantly increase. 
The reason for this expectation is that while more terms are being added, their weight is not increasing. 
As a consequence, these terms can also be fit into new partitions which reduces the error relative to an unpartitioned Hamiltonian. 
This implies our technique—and improvements made to the partition size—would continue to have a constant rate decrease in total Trotter error that is independent on array size.  
This is the exact behavior seen in the data of Figure~\ref{fig:scalability_proxy}.
For the electronic structure Hamiltonians, the fermion to qubit mapping used was the Bravyi-Kitaev mapping \cite{seeley2012bk}. 
The weight of Pauli terms increases logarithmically, so in this experiment, the expected behavior is a logarithmic curve. 
This is because logarithmically, terms are being added which cannot fit into our partition size (n=3,5). 
This however can be mitigated as techniques exist to have constant weight pauli terms \cite{derby2021compact}
We notice that there is noise in some of the ratios and we believe this is an artifact of the partitioning heuristics used. 
Overall, this proxy measurement gives evidence towards the scalability of constant-size partitions for reducing Trotter error as the ratio of non-commutation appears to have no dependence on quantum lattice size.

\section{Related Works \label{sec:relatedworks}} 
Trotterization error has been extensively studied, resulting in various strategies aimed at mitigating and managing these errors. Gui et al. \cite{gui2021termgroupingtravellingsalesperson} demonstrated that grouping neighboring terms in the Trotter step ordering can reduce errors by effectively clustering commuting operations. 
Additionally, a recent survey shows early work for rewriting certain partitions of the Hamiltonian to save on error per Trotter step \cite{Mart_nez_Mart_nez_2023}. However, these partitions are not general to all Hamiltonians of interest, and are also restricted to unitaries with special properties, making adaptability to input very difficult. 

Theoretical advancements, including higher-order Trotter decompositions \cite{Campbell_2019}, systematically eliminate specific-order errors through symmetric expansions. Our method can provide better performance due to the partial Trotter decomposition. By rewriting the non-commuting terms with minimal error, the error bound is reduced, which complements the optimizations and techniques described above in practice. 

Compiler optimizations for quantum Hamiltonian simulation, outside of error reduction,  have also been extensively studied.  Simultaneous diagonalization of commuting Pauli strings~\cite{Cowtan_2020, cowtan2020genericcompilationstrategyunitary, van_den_Berg_2020} is one early type of approach.
They are later outperformed by reordering-based gate cancellation~\cite{li2021paulihedralgeneralizedblockwisecompiler, gui2021termgroupingtravellingsalesperson, anastasiou2022tetrisadaptvqe} and Pauli network synthesis \cite{debrugiere2024fastershortersynthesishamiltonian,paykin2023pcoastpaulibasedquantumcircuit, schmitz2023graphoptimizationperspectivelowdepth}.

Similar work casted reordering and synthesis of the Trotter step as a travelling salesman graph problem, \cite{schmitz2023graphoptimizationperspectivelowdepth}  which was able to reduce the depth of Trotter steps substantially. 
Unlike our goal of grouping by commutation for merger across Trotter steps, the authors of this work framed Trotter reordering for optimization within a singular step.

The recent work QuCLEAR~\cite{liu2025quclearcliffordextractionabsorption} investigated extraction and absorption for Clifford gates in quantum Hamiltonian simulation, but it requires updating the observable. This work does not change other parts of the circuit, and the compiled Hamiltonian time evolution operator can be freely reused.
Moreover, all of them rely on the vanilla error bound of Trotterization and do not consider the fine-grained error scaling. 
Finally, other works focused on fermion to qubit mappings for cancellation of gates \cite{Liu_2025},\cite{Liu:2024vbw}
but this is specific to Fermionic Hamiltonians, and is complementary to our work due to the mapping of the Hamiltonian into the spin representation being assumed as input to Kernpiler. 


Unitary decomposition has been investigated mostly in a generic manner and separately from Hamiltonian mapping. 
Initial advancements, such as the quantum Shannon decomposition \cite{Shende_2006}, demonstrated how arbitrary unitaries can be decomposed into single- and two-qubit unitaries. Recent studies have precisely quantified the number of gates required for unitary operations, notably demonstrating that any 3-qubit unitary can be decomposed into a maximum of 19 CNOT gates \cite{PhysRevApplied.22.034019}. Although still above the theoretical minimum, these advances represent considerable progress. Additionally, numerical methods, while traditionally offering lower accuracy, provide intuitive trade-offs by significantly reducing gate counts, making them valuable for practical quantum computation applications \cite{Rakyta_2022}, \cite{bqskit}, \cite{Smith_2023}, \cite{Younis2020QFAST}. Overall, none of these general unitary decomposition methods take into account high level Hamiltonian information and therefore cannot adapt to high level structures of the unitary.

\section{Conclusion}
We introduced a novel compilation paradigm—leveraging partial Trotterization and strategic clustering of non-commuting Hamiltonian terms—that substantially improves the computational efficiency and accuracy of quantum Hamiltonian simulation. Reinforcement learning (via MCTS) proved effective in discovering optimized gate structures, and the RL search frequently identified recurring CNOT scaffolds and entanglement motifs; these learned circuit patterns suggest heuristic or graph-based synthesis algorithms that do not rely on RL, and motivate studying how CNOT scaffolds relate to accuracy convergence of overparameterized circuits.


\subsection*{Acknowledgements}
GL and ED were supported in part by the U.S. Department of Energy, Office of Science, Office of Advanced Scientific Computing Research through the Accelerated Research in Quantum Computing Program MACH-Q project., NSF CAREER Award No. CCF-2338773 and ExpandQISE Award No. OSI-2427020. GL is also supported by the Intel Rising Star Award. EM and EC were supported by the FY24 C2QA Postdoc Seed Funding Award from the Co-design Center for Quantum Advantage. EC was also supported in part by ARO MURI (award No.~SCON-00005095), and DoE (BNL contract No. 433702). EG was supported by the NASA Academic Mission Services, Contract No. NNA16BD14C and the Intelligent Systems Research and Development-3 (ISRDS-3) Contract 80ARC020D0010 under Co-design Center for Quantum Advantage (C$^2$QA) under Contract No. DE-SC0012704. AS acknowledges support from the U.S. Department of Energy, Office of Science, National Quantum Information Science Research Centers, Quantum Systems Accelerator.


\bibliographystyle{IEEEtranS}
\bibliography{apssamp}

@PREAMBLE{
 "\providecommand{\noopsort}[1]{}" 
 # "\providecommand{\singleletter}[1]{#1}%" 
}

@book{sutton2018reinforcement,
  title        = {Reinforcement Learning: An Introduction},
  author       = {Sutton, Richard S. and Barto, Andrew G.},
  year         = {2018},
  edition      = {2nd},
  publisher    = {MIT Press},
  address      = {Cambridge, MA},
  url          = {http://incompleteideas.net/book/the-book-2nd.html},
  isbn         = {9780262039246}
}

@misc{schmitz2023graphoptimizationperspectivelowdepth,
      title={Graph Optimization Perspective for Low-Depth Trotter-Suzuki Decomposition}, 
      author={Albert T. Schmitz and Nicolas P. D. Sawaya and Sonika Johri and A. Y. Matsuura},
      year={2023},
      eprint={2103.08602},
      archivePrefix={arXiv},
      primaryClass={quant-ph},
      url={https://arxiv.org/abs/2103.08602}, 
}

@article{lloyd1996universal,
  title={Universal quantum simulators},
  author={Lloyd, Seth},
  journal={Science},
  volume={273},
  number={5278},
  pages={1073--1078},
  year={1996},
  publisher={American Association for the Advancement of Science}
}

@article{seeley2012bk,
  title={The Bravyi--Kitaev transformation for quantum computation of electronic structure},
  author={Seeley, Jacob T. and Richard, Martin J. and Love, Peter J.},
  journal={The Journal of Chemical Physics},
  volume={137},
  number={22},
  pages={224109},
  year={2012}
}

@article{derby2021compact,
  title={Compact fermion-to-qubit mappings},
  author={Derby, Charles and Klassen, Joel and Bausch, Johannes and Cubitt, Toby},
  journal={Physical Review B},
  volume={104},
  number={3},
  pages={035118},
  year={2021}
}

@inproceedings{Stavenger:2022wzz,
    author = "Stavenger, Timothy J. and Crane, Eleanor and Smith, Kevin C. and Kang, Christopher T. and Girvin, Steven M. and Wiebe, Nathan",
    title = "{C2QA - Bosonic Qiskit}",
    booktitle = "{26th IEEE High Performance Extreme Computing}",
    eprint = "2209.11153",
    archivePrefix = "arXiv",
    primaryClass = "quant-ph",
    doi = "10.1109/HPEC55821.2022.9926318",
    month = "9",
    year = "2022"
}

@article{Smith_2023,
   title={LEAP: Scaling Numerical Optimization Based Synthesis Using an Incremental Approach},
   volume={4},
   ISSN={2643-6817},
   url={http://dx.doi.org/10.1145/3548693},
   DOI={10.1145/3548693},
   number={1},
   journal={ACM Transactions on Quantum Computing},
   publisher={Association for Computing Machinery (ACM)},
   author={Smith, Ethan and Davis, Marc Grau and Larson, Jeffrey and Younis, Ed and Oftelie, Lindsay Bassman and Lavrijsen, Wim and Iancu, Costin},
   year={2023},
   month=feb, pages={1–23} }

@article{Mart_nez_Mart_nez_2023,
   title={Assessment of various Hamiltonian partitionings for the electronic structure problem on a quantum computer using the Trotter approximation},
   volume={7},
   ISSN={2521-327X},
   url={http://dx.doi.org/10.22331/q-2023-08-16-1086},
   DOI={10.22331/q-2023-08-16-1086},
   journal={Quantum},
   publisher={Verein zur Forderung des Open Access Publizierens in den Quantenwissenschaften},
   author={Martínez-Martínez, Luis A. and Yen, Tzu-Ching and Izmaylov, Artur F.},
   year={2023},
   month=aug, pages={1086} }

@inproceedings{Liu_2025,
   title={HATT: Hamiltonian Adaptive Ternary Tree for Optimizing Fermion-to-Qubit Mapping},
   url={http://dx.doi.org/10.1109/HPCA61900.2025.00022},
   DOI={10.1109/hpca61900.2025.00022},
   booktitle={2025 IEEE International Symposium on High Performance Computer Architecture (HPCA)},
   publisher={IEEE},
   author={Liu, Yuhao and Yao, Kevin and Hong, Jonathan and Froustey, Julien and Rrapaj, Ermal and Iancull, Costin and Li, Gushu and Shi, Yunong},
   year={2025},
   month=mar, pages={143–157} }

@article{low2019,
  title = {Hamiltonian {{Simulation}} by {{Qubitization}}},
  author = {Low, Guang Hao and Chuang, Isaac L.},
  year = {2019},
  month = jul,
  journal = {Quantum},
  volume = {3},
  pages = {163},
  publisher = {Verein zur F{\"o}rderung des Open Access Publizierens in den Quantenwissenschaften},
  doi = {10.22331/q-2019-07-12-163}
}

@techreport{Younis2020QFAST,
  author      = {Younis, Ed and Sen, Koushik and Yelick, Katherine and Iancu, Costin},
  title       = {QFAST: Quantum Synthesis Using a Hierarchical Continuous Circuit Space},
  institution = {University of California, Berkeley, EECS Dept.},
  number      = {UCB/EECS-2020-53},
  year        = {2020},
  url         = {http://www2.eecs.berkeley.edu/Pubs/TechRpts/2020/EECS-2020-53.pdf}
}

@inproceedings{bqskit,
  title     = {BQSKit: The Berkeley Quantum Synthesis Toolkit},
  author    = {Nam, Yunseong and Ross, Nicholas J. and Su, Pei-Hsuan and Younis, Ed and Iancu, Costin C. and Lavrijsen, Wim and Brown, Kenneth R.},
  booktitle = {2020 IEEE International Conference on Quantum Computing and Engineering (QCE)},
  year      = {2020},
  pages     = {402--408},
  doi       = {10.1109/QCE49297.2020.00063}
}

@article{Childs_2021,
   title={Theory of Trotter Error with Commutator Scaling},
   volume={11},
   ISSN={2160-3308},
   url={http://dx.doi.org/10.1103/PhysRevX.11.011020},
   DOI={10.1103/physrevx.11.011020},
   number={1},
   journal={Physical Review X},
   publisher={American Physical Society (APS)},
   author={Childs, Andrew M. and Su, Yuan and Tran, Minh C. and Wiebe, Nathan and Zhu, Shuchen},
   year={2021},
   month=feb }

@misc{gui2021termgroupingtravellingsalesperson,
      title={Term Grouping and Travelling Salesperson for Digital Quantum Simulation}, 
      author={Kaiwen Gui and Teague Tomesh and Pranav Gokhale and Yunong Shi and Frederic T. Chong and Margaret Martonosi and Martin Suchara},
      year={2021},
      eprint={2001.05983},
      archivePrefix={arXiv},
      primaryClass={quant-ph},
      url={https://arxiv.org/abs/2001.05983}, 
}

@inbook{Hatano_2005,
   title={Finding Exponential Product Formulas of Higher Orders},
   ISBN={9783540315155},
   ISSN={1616-6361},
   url={http://dx.doi.org/10.1007/11526216_2},
   DOI={10.1007/11526216_2},
   booktitle={Quantum Annealing and Other Optimization Methods},
   publisher={Springer Berlin Heidelberg},
   author={Hatano, Naomichi and Suzuki, Masuo},
   year={2005},
   month=nov, pages={37–68} }

@article{Shende_2006,
   title={Synthesis of quantum-logic circuits},
   volume={25},
   ISSN={1937-4151},
   url={http://dx.doi.org/10.1109/TCAD.2005.855930},
   DOI={10.1109/tcad.2005.855930},
   number={6},
   journal={IEEE Transactions on Computer-Aided Design of Integrated Circuits and Systems},
   publisher={Institute of Electrical and Electronics Engineers (IEEE)},
   author={Shende, V.V. and Bullock, S.S. and Markov, I.L.},
   year={2006},
   month=jun, pages={1000–1010} }

@article{Kalajdzievski_2018,
   title={Continuous-variable gate decomposition for the Bose-Hubbard model},
   volume={97},
   ISSN={2469-9934},
   url={http://dx.doi.org/10.1103/PhysRevA.97.062311},
   DOI={10.1103/physreva.97.062311},
   number={6},
   journal={Physical Review A},
   publisher={American Physical Society (APS)},
   author={Kalajdzievski, Timjan and Weedbrook, Christian and Rebentrost, Patrick},
   year={2018},
   month=jun }

@misc{kaur2025optimizedquantumcircuitpartitioning,
      title={Optimized Quantum Circuit Partitioning Across Multiple Quantum Processors}, 
      author={Eneet Kaur and Hassan Shapourian and Jiapeng Zhao and Michael Kilzer and Ramana Kompella and Reza Nejabati},
      year={2025},
      eprint={2501.14947},
      archivePrefix={arXiv},
      primaryClass={quant-ph},
      url={https://arxiv.org/abs/2501.14947}, 
}

@article{Daei_2020,
   title={Optimized Quantum Circuit Partitioning},
   volume={59},
   ISSN={1572-9575},
   url={http://dx.doi.org/10.1007/s10773-020-04633-8},
   DOI={10.1007/s10773-020-04633-8},
   number={12},
   journal={International Journal of Theoretical Physics},
   publisher={Springer Science and Business Media LLC},
   author={Daei, Omid and Navi, Keivan and Zomorodi-Moghadam, Mariam},
   year={2020},
   month=nov, pages={3804–3820} }

@article{Childs_2019,
   title={Faster quantum simulation by randomization},
   volume={3},
   ISSN={2521-327X},
   url={http://dx.doi.org/10.22331/q-2019-09-02-182},
   DOI={10.22331/q-2019-09-02-182},
   journal={Quantum},
   publisher={Verein zur Forderung des Open Access Publizierens in den Quantenwissenschaften},
   author={Childs, Andrew M. and Ostrander, Aaron and Su, Yuan},
   year={2019},
   month=sep, pages={182} }

@article{PhysRevApplied.22.034019,
  title = {Beyond quantum Shannon decomposition: Circuit construction for $n$-qubit gates based on block-$ZXZ$ decomposition},
  author = {Krol, Anna M. and Al-Ars, Zaid},
  journal = {Phys. Rev. Appl.},
  volume = {22},
  issue = {3},
  pages = {034019},
  numpages = {7},
  year = {2024},
  month = {Sep},
  publisher = {American Physical Society},
  doi = {10.1103/PhysRevApplied.22.034019},
  url = {https://link.aps.org/doi/10.1103/PhysRevApplied.22.034019}
}

@article{Rakyta_2022,
   title={Approaching the theoretical limit in quantum gate decomposition},
   volume={6},
   ISSN={2521-327X},
   url={http://dx.doi.org/10.22331/q-2022-05-11-710},
   DOI={10.22331/q-2022-05-11-710},
   journal={Quantum},
   publisher={Verein zur Forderung des Open Access Publizierens in den Quantenwissenschaften},
   author={Rakyta, Péter and Zimborás, Zoltán},
   year={2022},
   month=may, pages={710} }

@misc{li2021paulihedralgeneralizedblockwisecompiler,
      title={Paulihedral: A Generalized Block-Wise Compiler Optimization Framework For Quantum Simulation Kernels}, 
      author={Gushu Li and Anbang Wu and Yunong Shi and Ali Javadi-Abhari and Yufei Ding and Yuan Xie},
      year={2021},
      eprint={2109.03371},
      archivePrefix={arXiv},
      primaryClass={quant-ph},
      url={https://arxiv.org/abs/2109.03371}, 
}

@article{Campbell_2019,
   title={Random Compiler for Fast Hamiltonian Simulation},
   volume={123},
   ISSN={1079-7114},
   url={http://dx.doi.org/10.1103/PhysRevLett.123.070503},
   DOI={10.1103/physrevlett.123.070503},
   number={7},
   journal={Physical Review Letters},
   publisher={American Physical Society (APS)},
   author={Campbell, Earl},
   year={2019},
   month=aug }

@software{cirq_developers_2024_11398048,
  author       = {Cirq Developers},
  title        = {Cirq},
  month        = may,
  year         = 2024,
  publisher    = {Zenodo},
  version      = {v1.4.0},
  doi          = {10.5281/zenodo.11398048},
  url          = {https://doi.org/10.5281/zenodo.11398048}
}

@misc{Liu:2024vbw,
    author = "Liu, Yuhao and Che, Shize and Zhou, Junyu and Shi, Yunong and Li, Gushu",
    title = "{Fermihedral: On the Optimal Compilation for Fermion-to-Qubit Encoding}",
    eprint = "2403.17794",
    archivePrefix = "arXiv",
    primaryClass = "quant-ph",
    doi = "10.1145/3620666.3651371",
    month = "3",
    year = "2024"
}

@article{Killoran_2019,
   title={Strawberry Fields: A Software Platform for Photonic Quantum Computing},
   volume={3},
   ISSN={2521-327X},
   url={http://dx.doi.org/10.22331/q-2019-03-11-129},
   DOI={10.22331/q-2019-03-11-129},
   journal={Quantum},
   publisher={Verein zur Forderung des Open Access Publizierens in den Quantenwissenschaften},
   author={Killoran, Nathan and Izaac, Josh and Quesada, Nicolás and Bergholm, Ville and Amy, Matthew and Weedbrook, Christian},
   year={2019},
   month=mar, pages={129} }

@misc{qiskit2024,
      title={Quantum computing with {Q}iskit},
      author={Javadi-Abhari, Ali and Treinish, Matthew and Krsulich, Kevin and Wood, Christopher J. and Lishman, Jake and Gacon, Julien and Martiel, Simon and Nation, Paul D. and Bishop, Lev S. and Cross, Andrew W. and Johnson, Blake R. and Gambetta, Jay M.},
      year={2024},
      doi={10.48550/arXiv.2405.08810},
      eprint={2405.08810},
      archivePrefix={arXiv},
      primaryClass={quant-ph}
}

@misc{mcclean2019openfermionelectronicstructurepackage,
      title={OpenFermion: The Electronic Structure Package for Quantum Computers}, 
      author={Jarrod R. McClean and Kevin J. Sung and Ian D. Kivlichan and Yudong Cao and Chengyu Dai and E. Schuyler Fried and Craig Gidney and Brendan Gimby and Pranav Gokhale and Thomas Häner and Tarini Hardikar and Vojtěch Havlíček and Oscar Higgott and Cupjin Huang and Josh Izaac and Zhang Jiang and Xinle Liu and Sam McArdle and Matthew Neeley and Thomas O'Brien and Bryan O'Gorman and Isil Ozfidan and Maxwell D. Radin and Jhonathan Romero and Nicholas Rubin and Nicolas P. D. Sawaya and Kanav Setia and Sukin Sim and Damian S. Steiger and Mark Steudtner and Qiming Sun and Wei Sun and Daochen Wang and Fang Zhang and Ryan Babbush},
      year={2019},
      eprint={1710.07629},
      archivePrefix={arXiv},
      primaryClass={quant-ph},
      url={https://arxiv.org/abs/1710.07629}, 
}

@misc{crane2024,
      title={Hybrid Oscillator-Qubit Quantum Processors: Simulating Fermions, Bosons, and Gauge Fields}, 
      author={Eleanor Crane and Kevin C. Smith and Teague Tomesh and Alec Eickbusch and John M. Martyn and Stefan Kühn and Lena Funcke and Michael Austin DeMarco and Isaac L. Chuang and Nathan Wiebe and Alexander Schuckert and Steven M. Girvin},
      year={2024},
      eprint={2409.03747},
      archivePrefix={arXiv},
      primaryClass={quant-ph},
      url={https://arxiv.org/abs/2409.03747}, 
}

@misc{dawson2005solovaykitaevalgorithm,
      title={The Solovay-Kitaev algorithm}, 
      author={Christopher M. Dawson and Michael A. Nielsen},
      year={2005},
      eprint={quant-ph/0505030},
      archivePrefix={arXiv},
      primaryClass={quant-ph},
      url={https://arxiv.org/abs/quant-ph/0505030}, 
}

@article{2018efficienttwirling,
	author = {Endo, Suguru and Benjamin, Simon C. and Li, Ying},
	doi = {10.1103/physrevx.8.031027},
	issn = {2160-3308},
	journal = {Physical Review X},
	month = {Jul},
	number = {3},
	publisher = {American Physical Society (APS)},
	title = {Practical Quantum Error Mitigation for Near-Future Applications},
	url = {http://dx.doi.org/10.1103/PhysRevX.8.031027},
	volume = {8},
	year = {2018}
}

@article{2013PhRvA..88a2314G,
	adsurl = {https://ui.adsabs.harvard.edu/abs/2013PhRvA..88a2314G},
	archiveprefix = {arXiv},
	author = {{Geller}, Michael R. and {Zhou}, Zhongyuan},
	doi = {10.1103/PhysRevA.88.012314},
	eid = {012314},
	eprint = {1305.2021},
	issn = {1094-1622},
	journal = {Physical Review A},
	month = jul,
	number = {1},
	pages = {012314},
	primaryclass = {quant-ph},
	publisher = {American Physical Society (APS)},
	title = {{Efficient error models for fault-tolerant architectures and the Pauli twirling approximation}},
	url = {http://dx.doi.org/10.1103/PhysRevA.88.012314},
	volume = {88},
	year = 2013
}

@misc{Winick:2022scr,
    author = "Winick, Adam and Wallman, Joel J. and Dahlen, Dar and Hincks, Ian and Ospadov, Egor and Emerson, Joseph",
    title = "{Concepts and conditions for error suppression through randomized compiling}",
    eprint = "2212.07500",
    archivePrefix = "arXiv",
    primaryClass = "quant-ph",
    month = "12",
    year = "2022"
}

@article{2016efficienttwirling,
	author = {Wallman, Joel J. and Emerson, Joseph},
	doi = {10.1103/physreva.94.052325},
	issn = {2469-9934},
	journal = {Physical Review A},
	month = {Nov},
	number = {5},
	publisher = {American Physical Society (APS)},
	title = {Noise tailoring for scalable quantum computation via randomized compiling},
	url = {http://dx.doi.org/10.1103/PhysRevA.94.052325},
	volume = {94},
	year = {2016}
}

@article{PhysRevA.94.052325,
  title = {Noise tailoring for scalable quantum computation via randomized compiling},
  author = {Wallman, Joel J. and Emerson, Joseph},
  journal = {Phys. Rev. A},
  volume = {94},
  issue = {5},
  pages = {052325},
  numpages = {9},
  year = {2016},
  month = {Nov},
  publisher = {American Physical Society},
  doi = {10.1103/PhysRevA.94.052325},
  url = {https://link.aps.org/doi/10.1103/PhysRevA.94.052325}
}

@book{nielsen2010quantum,
  title={Quantum computation and quantum information},
  author={Nielsen, Michael A and Chuang, Isaac L},
  year={2010},
  publisher={Cambridge university press}
}

@article{Cao_2019,
	doi = {10.1021/acs.chemrev.8b00803},
  
	url = {https://doi.org/10.1021%2Facs.chemrev.8b00803},
  
	year = 2019,
	month = {aug},
  
	publisher = {American Chemical Society ({ACS})},
  
	volume = {119},
  
	number = {19},
  
	pages = {10856--10915},
  
	author = {Yudong Cao and Jonathan Romero and Jonathan P. Olson and Matthias Degroote and Peter D. Johnson and M{\'{a}
}ria Kieferov{\'{a}} and Ian D. Kivlichan and Tim Menke and Borja Peropadre and Nicolas P. D. Sawaya and Sukin Sim and Libor Veis and Al{\'{a}}n Aspuru-Guzik},
  
	title = {Quantum Chemistry in the Age of Quantum Computing},
  
	journal = {Chemical Reviews}
}

@misc{anastasiou2022tetrisadaptvqe,
      title={TETRIS-ADAPT-VQE: An adaptive algorithm that yields shallower, denser circuit ans\"atze}, 
      author={Panagiotis G. Anastasiou and Yanzhu Chen and Nicholas J. Mayhall and Edwin Barnes and Sophia E. Economou},
      year={2022},
      eprint={2209.10562},
      archivePrefix={arXiv},
      primaryClass={quant-ph}
}

@article{Bauer:2023qgm,
    author = "Bauer, Christian W. and Davoudi, Zohreh and Klco, Natalie and Savage, Martin J.",
    title = "{Quantum simulation of fundamental particles and forces}",
    eprint = "2404.06298",
    archivePrefix = "arXiv",
    primaryClass = "hep-ph",
    reportNumber = "IQuS@UW-21-052",
    doi = "10.1038/s42254-023-00599-8",
    journal = "Nature Rev. Phys.",
    volume = "5",
    number = "7",
    pages = "420--432",
    year = "2023"
}

@article{Childs2012,
    title = {Hamiltonian simulation using linear combinations of unitary operations},
    volume = {12},
    issn = {1533-7146},
    url = {http://dx.doi.org/10.26421/QIC12.11-12-1},
    doi = {10.26421/qic12.11-12-1},
    number = {11 \& 12},
    journal = {Quantum Information and Computation},
    author = {Childs, Andrew M. and Wiebe, Nathan},
    month = nov,
    year = {2012},
    note = {Publisher: Rinton Press},
    pages = {901--924},
}

@article{babbush_low-depth_2018,
    title = {Low-{Depth} {Quantum} {Simulation} of {Materials}},
    volume = {8},
    issn = {2160-3308},
    url = {https://link.aps.org/doi/10.1103/PhysRevX.8.011044},
    doi = {10.1103/PhysRevX.8.011044},
    language = {en},
    number = {1},
    urldate = {2025-02-28},
    journal = {Physical Review X},
    author = {Babbush, Ryan and Wiebe, Nathan and McClean, Jarrod and McClain, James and Neven, Hartmut and Chan, Garnet Kin-Lic},
    month = mar,
    year = {2018},
    pages = {011044},
}

@article{hemery_measuring_2024,
    title = {Measuring the {Loschmidt} {Amplitude} for {Finite}-{Energy} {Properties} of the {Fermi}-{Hubbard} {Model} on an {Ion}-{Trap} {Quantum} {Computer}},
    volume = {5},
    url = {https://link.aps.org/doi/10.1103/PRXQuantum.5.030323},
    doi = {10.1103/PRXQuantum.5.030323},
    number = {3},
    urldate = {2025-01-28},
    journal = {PRX Quantum},
    author = {Hémery, Kévin and Ghanem, Khaldoon and Crane, Eleanor and Campbell, Sara L. and Dreiling, Joan M. and Figgatt, Caroline and Foltz, Cameron and Gaebler, John P. and Johansen, Jacob and Mills, Michael and Moses, Steven A. and Pino, Juan M. and Ransford, Anthony and Rowe, Mary and Siegfried, Peter and Stutz, Russell P. and Dreyer, Henrik and Schuckert, Alexander and Nigmatullin, Ramil},
    month = aug,
    year = {2024},
    note = {Publisher: American Physical Society},
    pages = {030323},
}

@misc{liu2025quclearcliffordextractionabsorption,
      title={QuCLEAR: Clifford Extraction and Absorption for Quantum Circuit Optimization}, 
      author={Ji Liu and Alvin Gonzales and Benchen Huang and Zain Hamid Saleem and Paul Hovland},
      year={2025},
      eprint={2408.13316},
      archivePrefix={arXiv},
      primaryClass={quant-ph},
      url={https://arxiv.org/abs/2408.13316}, 
}

@article{Cowtan_2020,
   title={Phase Gadget Synthesis for Shallow Circuits},
   volume={318},
   ISSN={2075-2180},
   url={http://dx.doi.org/10.4204/EPTCS.318.13},
   DOI={10.4204/eptcs.318.13},
   journal={Electronic Proceedings in Theoretical Computer Science},
   publisher={Open Publishing Association},
   author={Cowtan, Alexander and Dilkes, Silas and Duncan, Ross and Simmons, Will and Sivarajah, Seyon},
   year={2020},
   month=may, pages={213–228} }

@misc{cowtan2020genericcompilationstrategyunitary,
      title={A Generic Compilation Strategy for the Unitary Coupled Cluster Ansatz}, 
      author={Alexander Cowtan and Will Simmons and Ross Duncan},
      year={2020},
      eprint={2007.10515},
      archivePrefix={arXiv},
      primaryClass={quant-ph},
      url={https://arxiv.org/abs/2007.10515}, 
}

@article{van_den_Berg_2020,
   title={Circuit optimization of Hamiltonian simulation by simultaneous diagonalization of Pauli clusters},
   volume={4},
   ISSN={2521-327X},
   url={http://dx.doi.org/10.22331/q-2020-09-12-322},
   DOI={10.22331/q-2020-09-12-322},
   journal={Quantum},
   publisher={Verein zur Forderung des Open Access Publizierens in den Quantenwissenschaften},
   author={van den Berg, Ewout and Temme, Kristan},
   year={2020},
   month=sep, pages={322} }

@misc{paykin2023pcoastpaulibasedquantumcircuit,
      title={PCOAST: A Pauli-based Quantum Circuit Optimization Framework}, 
      author={Jennifer Paykin and Albert T. Schmitz and Mohannad Ibrahim and Xin-Chuan Wu and A. Y. Matsuura},
      year={2023},
      eprint={2305.10966},
      archivePrefix={arXiv},
      primaryClass={quant-ph},
      url={https://arxiv.org/abs/2305.10966}, 
}

@misc{debrugiere2024fastershortersynthesishamiltonian,
      title={Faster and shorter synthesis of Hamiltonian simulation circuits}, 
      author={Timothee Goubault de Brugiere and Simon Martiel},
      year={2024},
      eprint={2404.03280},
      archivePrefix={arXiv},
      primaryClass={quant-ph},
      url={https://arxiv.org/abs/2404.03280}, 
}

\end{document}